\documentclass[aps,prd,preprint,superscriptaddress,preprintnumbers,eqsecnum,nofootinbib,nobibnotes,showpacs]{revtex4}

\usepackage{amsfonts,amssymb,amsmath}
\usepackage{graphicx}

\newcommand{\be}{\begin{equation}}
\newcommand{\bea}{\begin{eqnarray}}
\newcommand{\ee}{\end{equation}}
\newcommand{\eea}{\end{eqnarray}}
\newcommand{\bpi}{\begin{picture}}
\newcommand{\bce}{\begin{center}}
\newcommand{\epi}{\end{picture}}
\newcommand{\ece}{\end{center}}

\newcommand{\de}{{\Delta}_{\rm \mbox{\tiny E}}}  
\newcommand{\da}{\rm \mbox{\tiny E}}  

\newcommand{\sla}{\slash \hspace{-0.22cm}}

\begin{document}

\title{Chiral symmetry breaking with lattice propagators}

\author{A.~C. Aguilar}
\email{Arlene.Aguilar@ufabc.edu.br}
\affiliation{Federal University of ABC, CCNH, \\
Rua Santa Ad\'{e}lia 166, CEP 09210-170, Santo Andr\'{e}, Brazil.}

\author{J. Papavassiliou}
\email{Joannis.Papavassiliou@uv.es}
\affiliation{Department of Theoretical Physics and IFIC, 
University of Valencia-CSIC,
E-46100, Valencia, Spain.}

\begin{abstract}

We  study chiral  symmetry breaking  using the  standard  gap equation,
supplemented  with  the  infrared-finite  gluon propagator  and  ghost
dressing function obtained  from large-volume lattice simulations. One
of the most important ingredients  of this analysis is the non-abelian
quark-gluon  vertex, which controls  the way  the ghost  sector enters
into  the  gap  equation.   Specifically,  this  vertex  introduces  a
numerically crucial dependence on  the ghost dressing function and 
the quark-ghost scattering amplitude.  This latter quantity 
satisfies its own, previously unexplored, dynamical equation,
which  may be decomposed  into individual  integral equations  for its
various  form  factors.  In  particular,  the  scalar  form factor  is
obtained from an approximate version of the ``one-loop dressed'' integral
equation,   and  its  numerical   impact  turns   out  to   be  rather
considerable.  The  detailed numerical  analysis of the  resulting gap
equation reveals that the constituent quark mass obtained is about 300
MeV, while  fermions in the adjoint representation  acquire a mass
in the range of (750-962) MeV.

\end{abstract}


\maketitle

\section{Introduction}

The dynamical mechanism responsible for chiral symmetry breaking (CSB)
in QCD has  been the focal point of  extensive research during several
years~\cite{Politzer:1976tv,Miransky:1981rt,Atkinson:1988mw, Brown:1988bm, 
Williams:1989tv,Papavassiliou:1991hx,Haeri:1990yj,Hawes:1993ef,Roberts:1994dr,
Natale:1996eu,Fischer:2003rp,Aguilar:2005sb,Bowman:2005vx,Sauli:2006ba,Cornwall:2008da}.  The study  
of CSB in the continuum  involves almost invariably
some  version of  the  Schwinger-Dyson equation  (SDE)  for the  quark
propagator (gap  equation).  This  non-linear integral equation  has a
notoriously rich  structure, being extremely sensitive  to the details
of its kernel; the latter  is composed by the various non-perturbative
ingredients  entering into the  gap equation,  most notably  the gluon
propagator  and the  quark gluon  vertex. As is well-known,  the gap
equation  displays  ``critical''  behavior: the  support  of  the  kernel
throughout  the entire  range  of integration  must  exceed a  certain
critical  value in  order to  generate non-trivial  solutions  for the
quark propagator~\cite{Atkinson:1988mw}. Given that most of this support 
originates from the infrared region, i.e. around the QCD  mass scale of a 
few hundred MeV, the study of CSB through the gap equation furnishes
stringent probes on the various methods and models aiming towards a 
quantitative description of the non-perturbative sector of QCD.


In recent years, a large number of independent large-volume lattice 
simulations have furnished highly non-trivial information on 
the infrared (IR) behavior of two fundamental ingredients of 
pure Yang-Mills theories, namely the (quenched) gluon and ghost
propagators, for both $SU(2)$ and 
$SU(3)$~\cite{Cucchieri:2007md,Cucchieri:2009zt,Bogolubsky:2007ud,Bogolubsky:2009dc,Oliveira:2009eh,Bowman:2007du}. 
In particular, these simulations have firmly established that (in the Landau gauge) 
the QCD gluon propagator and the ghost dressing function are IR finite and 
non-vanishing~\cite{Aguilar:2008xm,Boucaud:2008ji,Dudal:2007cw}. 

Given that the lattice is expected to capture reliably the 
full non-perturbative information contained in the gluon and
ghost propagators, it is natural to explore their
consequences for CSB. To that end, in this article we use
lattice results for these Green's functions as inputs for the gap 
equation, and study the emerging CSB pattern for quarks (fundamental representation) and 
for fermions in the adjoint representation.
Specifically, we will employ the lattice data of~\cite{Bogolubsky:2007ud}, 
given that they perform $SU(3)$ simulations for both the gluon and the ghost propagators.  
As it will become clear from the results presented in the main body of the paper, 
the analysis carried out here may be regarded as a serious test 
of the robustness of the aforementioned lattice results, and can serve as  
a characteristic example of the rich phenomenology that one may extract with them.


The detailed implementation of the idea described above 
is far from straightforward, mainly due to the complicated structure 
of the gap equation, which makes it difficult to determine its exact dependence on the 
aforementioned lattice ingredients (propagators). 
In fact, of particular importance for the self-consistency of the 
whole picture is the role played by the ghost 
sector (see, e.g.~\cite{Fischer:2003rp}, and references therein). 
The way the ghost sector enters into the gap equation is through 
the fully-dressed quark-gluon vertex. 
Specifically, recall that, in virtually all treatments,  
the fully-dressed quark-gluon vertex is not obtained from the corresponding 
dynamical equation (the SDE of the vertex), but is rather expressed in terms of 
the quark propagator, such that  its Slavnov-Taylor identity (STI) is automatically  satisfied
(this procedure is known as the ``gauge technique''~\cite{Salam:1963sa}).
The STI itself contains explicit reference to both the ghost dressing function and the 
so-called ``quark-ghost scattering kernel''~\cite{Marciano:1977su}; the latter is given by its own 
dynamical equation, and, as we will see, its numerical impact to the solutions 
obtained from the gap equation is quite important.  
If, instead, one were to ``abelianize''  this  part  of  the
problem  by  assuming  that  the quark-gluon vertex satisfies
a  QED-like Ward identity rather than the correct (non-abelian) STI, the resulting
gap equation would contain the gluon propagator as its sole ingredient, a fact that 
would lead   to   an  apparent incompatibility,  in the  sense that  the kernel 
would not  exceed the critical  value (no CSB), 
or would fail to generate  realistic constituent
quark  masses~\cite{Papavassiliou:1991hx,Haeri:1990yj,Natale:1996eu,Cornwall:2008da}.

The main results of the present article may be summarized as follows:

({\bf a}) The non-abelian Ansatz for the 
full quark-gluon vertex depends explicitly 
on the the ghost-dressing function and the quark-ghost scattering kernel;
the latter quantity has a rather complicated Dirac
structure~\cite{Davydychev:2000rt}, being composed 
by four independent form factors [see Eq.~(\ref{Xi})]. As a consequence, the 
corresponding expressions for the form factors appearing in the Lorentz decomposition 
for the longitudinal part of the quark-gluon vertex [see Eq.~(\ref{Li})] 
are modified (with respect to the case where the  quark-ghost scattering kernel
is set to its tree-level value); their full form is presented in  Eq.~(\ref{expLi}).
As a result, the gap equation acquires a more complicated structure, 
given in Eqs.~(\ref{gpa})-(\ref{gpb}).  
To the best of our knowledge, both Eq.~(\ref{expLi}) and Eqs.~(\ref{gpa})-(\ref{gpb})
appear for the first time in the literature.

({\bf b}) The quark-ghost scattering kernel satisfies its own dynamical equation, 
which may be decomposed into individual integral equations for the various  
form factors entering into the gap equation. In turn, these integral equations depend, 
among other quantities, on the quark propagator, a fact which converts 
the full treatment of the problem into the solution of a complicated system of various
coupled integral equations. 

({\bf c}) In order to make the above system of equations more tractable, without
compromising its main features, we retain only the dependence of the gap 
equation on the scalar form factor
of the  quark-ghost scattering kernel, discarding all other form factors.
In addition, we choose a very particular kinematic configuration, 
which further simplifies the corresponding integral equation that 
determines the aforementioned quantity. The final   
``one-loop dressed'' equation is given in Eq.~(\ref{X0LD1}), 
and constitutes, to the best of our knowledge, a novel result.   
For the actual calculation of the scalar form factor 
we will use on the rhs of Eq.~(\ref{X0LD1})  the lattice results   
of~\cite{Bogolubsky:2007ud}, and then substitute the result 
(shown in Fig.~\ref{sk}) into the gap equation.  

({\bf d})
A well-known endemic shortcoming of all 
approaches based on the gauge-technique 
is that the transverse (i.e. identically conserved) 
part of the  (quark-gluon) vertex remains largely  
undetermined; this fact, in turn, distorts the 
cancellations of overlapping divergences, the 
multiplicative renormalizability of the Green's functions in 
question, and their compliance with the renormalization group (RG). 
The construction of the appropriate transverse piece has been carried out in detail 
for the case of QED~\cite{Kizilersu:2009kg}, but no real progress has been made in a non-abelian context. 
As is common practice, 
the aforementioned problem is remedied by  
multiplying (by hand) the kernel of the gap equation by 
an appropriate functions, which restores the desired properties.  
For the case at hand, the simplest quantity that accounts for the missing dynamics  
is the full ghost dressing-function. 
As we will explain in the corresponding subsection, 
this choice is dictated by the STI satisfied by the quark-gluon vertex,
and enforces the correct RG behavior of the dynamical (running) 
mass obtained from the gap equation. It should be stressed that 
the inclusion of the dressing-function has a considerable numerical impact 
on the obtained CSB solution, boosting up the quark mass to phenomenologically 
acceptable values.  

({\bf e}) 
After substituting all necessary ingredients comprising its kernel, 
the resulting gap equation is finally solved numerically, 
for two different cases. First, we study fermions in the fundamental 
representation (quarks), obtaining a  
quark mass that in the IR is about 300 MeV.
Second, we consider CSB with fermions in the adjoint representation; the 
latter are particularly interesting, 
due to the clear separation between 
chiral symmetry restoration and deconfinement they display~\cite{Karsch:1998qj,Basile:2005hw,Engels:2005rr}.
The corresponding mass obtained for the adjoint fermions is within the range  
(750-962) MeV, depending on the details of the quark-gluon vertex used.
This values are not too far from what one would expect naively, 
given the enhancement of 9/4 produced to the kernel of the adjoint gap equation 
due to the ratio of the Casimir eigenvalues
of the two representations.

The article is organized as follows. In Section~\ref{gap} 
we introduce the necessary notation, and review the 
general the structure of the gap equation. 
In Section~\ref{ghostgap} we first construct 
an Ansatz
for the quark-gluon vertex which makes full reference on the 
quark-ghost scattering kernel, and use this vertex to 
derive the corresponding gap equation.  Then, the ``one-loop dressed''  
approximation for the scalar part 
of the quark-ghost scattering kernel is set up, 
and the improvements necessary for restoring  
the correct RG properties are discussed in detail.  
In Section~\ref{numan} we present the main results of this work. In particular, 
after briefly reviewing the recent lattice results on the gluon and ghost propagators 
that enter into the gap equation, 
we proceed to the numerical solution of the gap equation. 
In Section~\ref{concl} we discuss our results 
and comment on possible future directions. Finally, in an Appendix we analyze for completeness 
the structure of the gap equation within the framework of the pinch technique (PT)~\cite{Cornwall:1982zr,Cornwall:1989gv,Binosi:2002ft}  
or, equivalently, the background field method (BFM)~\cite{Abbott:1980hw}.

\section{General structure of the gap equation}
\label{gap}

In this section we will introduce the basic definitions and ingredients necessary for
the study of the quark SDE (gap equation).  Then, we will give a special emphasis in the
construction of a general Ansatz for full fermion-gluon vertex, where its non-abelian character
will be kept intact. Finally, we will incorporate it into the gap equation and explore 
its effects. 
   
\subsection{Definitions and ingredients}

Let us first introduce the necessary notation.
In covariant gauges, the inverse of the full quark propagator in the Minkowski space  
has the general form~\cite{Marciano:1977su}
\be
S^{-1}(p) = \sla{p} -m -\Sigma(p)\,,  
\label{qprop}
\ee
where $m$  is the bare current quark mass, and  $\Sigma(p)$  the quark
self-energy. 
It is common practice to decompose    
$\Sigma(p)$, in terms of a Dirac vector component, $A(p^2)$, and a scalar
component, $B(p^2)$, which allow us to define the dynamical quark mass function as being the ratio   
\mbox{${\mathcal{M}}(p^2)= B(p^2)/A(p^2)$}, explicitly we have~\cite{Roberts:1994dr}
\be
S^{-1}(p) =  A(p^2)\,\sla{p} - B(p^2) \mathbb{I} 
= A(p^2)[\sla{p}-{\mathcal{M}}(p^2) \mathbb{I}] \,,
\label{qpropAB}
\ee
where $\mathbb{I}$ is the identity matrix, and the 
term \mbox{$A^{-1}(p^2)$} is often referred to in the literature  as the ``fermion wave function''.
Note that, the fermion acquires a dynamical mass as long as $B(p^2)$ is different from zero. Therefore,
the CSB will be signaled when we obtain $B(p^2) \neq 0$.

%
\begin{figure}[ht]
\includegraphics[scale=0.6]{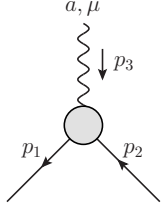}
\caption{The full fermion-gluon vertex.}
\label{fig_vqg}
\end{figure}

In addition, the gluon propagator $\Delta_{\mu\nu}(q)$ in the covariant  renormalizable ($R_\xi$) gauges, 
has the form
\be \Delta_{\mu\nu}(q)=-i\left[ P_{\mu\nu}(q)\Delta(q^2) +\xi\frac{q_\mu q_\nu}{q^4}\right],
\label{prop_cov}
\ee
where $\xi$ denotes the gauge-fixing parameter, and 
\be
P_{\mu\nu}(q)= g_{\mu\nu} - q_\mu q_\nu /q^2 \,,
\ee
is the usual transverse projector.  In this work, we are particularly interested in the Landau 
gauge which is reached when $\xi=0$. Moreover, the full ghost propagator $D(q^2)$ and its dressing function $F(q^2)$
are related by 
\be
D(q^2)= \frac{iF(q^2)}{q^2}.
\label{ghostdress}
\ee

An essential ingredient in our study is the fermion-gluon vertex, represented in Fig.~\ref{fig_vqg}, and given by  
\be
\Gamma_{\mu}^{a}(p_1,p_2,p_3) = g T^{a} \Gamma_{\mu}(p_1,p_2,p_3) \,,
\ee
with $T^{a}$ ($a =1,2, ..., N^2-1$) being the generators of the group  $SU(N)$  
where the fermions are assigned. The matrices $T^{a}$ 
are hermitian and traceless, generating the closed 
algebra
\be
[T^{a}, T^{b}]  = i f^{abc} T^{c} \,,
\label{comrel}
\ee
where $f^{abc}$ are the (totally antisymmetric) structure constants. 
In the case of $SU(3)$, and for fermions in the fundamental representations (quarks),  
we have that $T^{a} = \lambda^{a}/2$, where $\lambda^{a}$ are the Gell-Mann  matrices.
When fermions are in the adjoint, $(T^{a})^{bc} = -if^{abc}$. 

\begin{center}
\begin{figure}[t]
\includegraphics[scale=0.8]{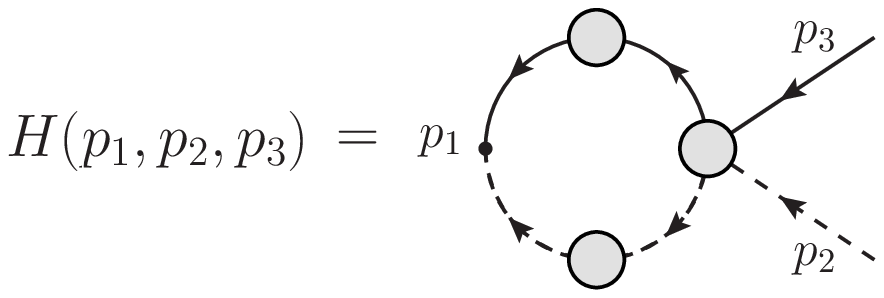}
\caption{Diagrammatic representation of the fermion-ghost scattering kernel $H(p_1,p_2,p_3)$.}
\label{fig_sk}
\end{figure}
\end{center}

The vertex  $\Gamma_{\mu}(p_1,p_2,p_3)$  satisfies the fundamental STI~\cite{Marciano:1977su}
\be
p_3^{\mu}\Gamma_{\mu}(p_1,p_2,p_3) = 
F(p_3)[S^{-1}(-p_1) H(p_1,p_2,p_3) - {\overline H}(p_2,p_1,p_3) S^{-1}(p_2)]\,,
\label{STI}
\ee
where the fermion-ghost scattering kernel  $H(p_1,p_2,p_3)$ is defined diagrammatically in Fig.~\ref{fig_sk}, and
is written as
\be
H^{a}(p_1,p_2,p_3) = T^{a} H(p_1,p_2,p_3) \,.
\ee

The kernel $H(p_1,p_2,p_3)$ and the 
``conjugated'' ${\overline H}(p_2,p_1,p_3)$ have the 
following Lorentz decomposition~\cite{Davydychev:2000rt} (note the change 
$p_1 \leftrightarrow p_2$ in the arguments of the latter)
\bea
H(p_1,p_2,p_3) &=& X_0 \mathbb{I}  
+X_1 \sla{p_1} +  
X_2  \sla{p_2} +
X_3 \tilde\sigma_{\mu\nu}p_1^{\mu} p_2^{\nu} \,,
\nonumber\\ 
{\overline H}(p_2,p_1,p_3) &=& 
{\overline X}_0 \mathbb{I} 
-{\overline X}_2 \sla{p_1}  
-{\overline X}_1 \sla{p_2} 
+{\overline X}_3 \tilde\sigma_{\mu\nu}p_1^{\mu} p_2^{\nu} \,,
\label{Xi}
\eea
where the form factors $X_i$ are functions of the momenta,
$X_i=X_i(p_1,p_2,p_3)$,  
and we use the notation 
${\overline X}_i (p,r,q) \equiv X_i (r,p,q)$
and $\tilde\sigma_{\mu\nu} \equiv \frac{1}{2}[\gamma_{\mu},\gamma_{\nu}]$
(Note the difference 
between $\tilde\sigma_{\mu\nu}$ and 
the usually defined $\sigma_{\mu\nu} = \frac{i}{2}[\gamma_{\mu},\gamma_{\nu}]$).
    
\subsection{The renormalized gap equation}

\begin{figure}[ht]
\includegraphics[scale=0.6]{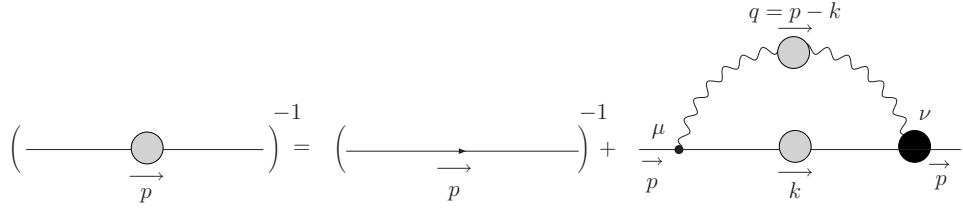}
\caption{ The SDE for the fermion  propagator given by Eq.~(\ref{senergy}). The gray
blobs represent the fully dressed gluon and quark propagators, while the black blob denotes the dressed 
fermion-gluon vertex.}
\label{self}
\end{figure}

The SDE for the fermion propagator is diagrammatically represented
in Fig.~\ref{self}. Using the momenta flow and Lorentz indices indicated in 
Fig.~\ref{self}, the gap equation can be written as

%
%
\begin{equation}
S^{-1}(p)= \sla{p} -m -C_{\rm r}g^2\int_k
\Gamma_{\mu}^{[0]}S(k)\Gamma_{\nu}(-p,k,q)\Delta^{\mu\nu}(q) \,,
\label{senergy}
\end{equation}
where $q\equiv p-k$, \mbox{$\int_{k}\equiv\mu^{2\varepsilon}(2\pi)^{-d}\int\!d^d k$}, 
with $d=4-\epsilon$ the dimension of space-time, and $\Gamma_{\mu}^{[0]}$ is the fermion-gluon vertex at tree level. 
$C_{\rm r}$  is the Casimir eigenvalue of the given fermion representation  
($r={\rm F}$  for the fundamental, and $r{=\rm A}$ for the  adjoint).  
More specifically, for the gauge group  $SU(3)$, we have \mbox{$C_{\rm A}=3$} and \mbox{$C_{\rm F}=4/3$}. 
Note that, $m$ is a current fermion mass, the same appearing in the QCD Lagrangian, and  
in case  of $m\neq 0$ in Eq.~(\ref{senergy}), the chiral symmetry is explicitly broken.
In this work, we will consider the case $m=0$, i.e.,  
the chiral symmetry is kept intact at the Langragian level.

All quantities appearing in Eq.~(\ref{senergy}) are
unrenormalized; they are related 
to their respective renormalized counterparts, denoted
with a subscript ``R'', through the relations~\cite{Itzykson:rh} 
\bea
S_R(p ; \mu)&=& Z_F^{-1}(\mu)S(p)\,, \nonumber\\
\Delta_R(q; \mu) &=& Z_{A}^{-1}(\mu) \Delta(q)\,,
\nonumber\\
F_R(q ; \mu) &=& Z_{c}^{-1}(\mu) F(q)\,,
\nonumber\\ 
\Gamma^{\nu}_R(p,k,q;\mu)  &=& Z_{1}(\mu)\Gamma^{\nu}(p,k,q)\,,
\nonumber\\
g_R(\mu) &=& Z_g^{-1}(\mu) g\, =  Z_1^{-1} Z_F^{1} Z_A^{1/2}g\,,
\label{renconst}
\eea
where $Z_F$, $Z_{A}$, $Z_{c}$,  $Z_{1}$, and $Z_g$ are the corresponding
renormalization constants.
Substituting Eqs.~(\ref{renconst}) 
into Eq.~(\ref{senergy}), we obtain
\begin{equation}
S^{-1}_{R}(p)= Z_F\sla{p} -Z_{1}C_{\rm r}g_R^2\int_k\,
\gamma_{\mu}S_R(k)\Gamma_{R\,\nu}(-p,k,q)\Delta^{\mu\nu}_R(q) \,.
\label{rsenergy}
\end{equation}
In addition, the STIs of Eq.~(\ref{STI}) imposes  
the all-order constraint 
\begin{equation}
Z_1= Z_c^{-1} Z_F Z_H^{-1} \,,
\end{equation}
where $Z_H$ is the renormalization constant needed for the quark-ghost kernel, i.e., 
\mbox{$H = Z_H^{-1} H_{R}$}. 
Now, in the Landau gauge, both  the quark self-energy and the quark-ghost kernel are 
finite at one-loop; thus, at that order, $Z_F = Z_H =1$, and, therefore,   
$Z_1= Z_c^{-1}$, i.e.,  at one-loop  the quark-gluon vertex renormalizes as the inverse  
the ghost propagator.
Imposing the above approximation in the  Eq.~(\ref{rsenergy}), we obtain
\begin{equation}
S^{-1}(p)= \sla{p} - Z_c^{-1} C_{\rm r}g^2\int_k\,
\gamma_{\mu}S(k)\Gamma_{\nu}(-p,k,q)\Delta^{\mu\nu}(q) \,,
\label{rsenergy1}
\end{equation}
where we have suppressed the subscript ``R'' to avoid notation clutter.


\section{Influence of the ghost sector on the gap equation}
\label{ghostgap}

In this rather lengthy and technical section we study in detail how the 
ghost sector enters into the gap equation. To that end, in subsection~\ref{ss1} we will 
use the STI to determine the dependence of the  
form factors of the longitudinal part of the quark-gluon vertex
on the corresponding form factors $X_i$ appearing in the Dirac decomposition of the 
quark-ghost scattering kernel, given in Eq.~(\ref{Xi}). Then, we will use the resulting vertex 
in order to derive the most general expression for the gap equation, 
displaying the dependence on all form factors  $X_i$.  
In subsection~\ref{ss2} we derive the ``one-loop dressed'' expression for the scalar 
form factor $X_0$, which is the only one that will be considered in the ensuing analysis. 
The modifications introduced into some standard forms of the quark-gluon vertex vertex, 
and the form of the gap equations obtained with them are presented in subsection \ref{ss3}. 
Finally, the adjustments necessary in order to enforce the 
correct RG properties of the gap equation are discussed in subsection~\ref{ss4}.

\subsection{The full fermion-gluon vertex}
\label{ss1}

The most general Lorentz decomposition for the longitudinal part of the vertex $\Gamma_{\mu}(p_1,p_2,p_3)$
can be written as~\cite{Davydychev:2000rt}
\be
\Gamma_{\mu}(p_1,p_2,p_3) = 
  L_1 \gamma_{\mu}
+ L_2 (\sla{p_1} - \sla{p_2})(p_1-p_2)_{\mu} 
+ L_3 (p_1-p_2)_{\mu} 
+ L_4 \tilde\sigma_{\mu\nu}(p_1-p_2)^{\nu} \,,
\label{Li}
\ee
where $L_i$ are the form factors, whose  dependence on the momenta 
has been suppressed, in order to keep a compact notation {\it i.e.} $L_i=L_i(p_1,p_2,p_3)$. The tree level expression is 
recovered setting $L_1=1$ and $L_2=L_3=L_4=0$; then, 
$\Gamma_{\mu}^{[0]}(p_1,p_2,p_3) = \gamma_{\mu}$.

Due to the fact that the behavior of the vertex $\Gamma_{\mu}(p_1,p_2,p_3)$ is constrained
by the STI of Eq.~(\ref{STI}), the form factors $L_i$'s appearing into
the Eq.~(\ref{Li}) will be given in terms of the form factors $X_i$'s of Eq.~(\ref{Xi}).

More specifically, using the standard decomposition of $S^{-1}(p)$ expressed in Eq.~(\ref{qpropAB}), 
it is relatively straightforward to demonstrate that the rhs of Eq.~(\ref{STI}) becomes 
\be
p_3^{\mu}\Gamma_{\mu}(p_1,p_2,p_3) = 
F(p_3) [C_0 \mathbb{I} 
+C_1 \sla{p_1} +  
C_2  \sla{p_2} +
C_3 \tilde\sigma_{\mu\nu}p_1^{\mu} p_2^{\nu}] \,,
\label{VCi}
\ee
with 
\bea
C_0  &=& - A(p_1)\left(p_1^2 X_1  + p_1\!\cdot\!p_2 X_2 \right) + 
A(p_2)\left(p_2^2 {\overline X}_1 + p_1\!\cdot\!p_2 {\overline X}_2\right) 
- B(p_1) X_0 - B(p_2){\overline X}_0 \,;
\nonumber\\ 
C_1  &=& A(p_1)\left(p_1\!\cdot\!p_2 X_3 -X_0 \right) - p_2^2 A(p_2){\overline X}_3
- B(p_1) X_1 + B(p_2){\overline X}_2 \,; 
\nonumber\\ 
C_2  &=&  A(p_2)\left(p_1\!\cdot\!p_2 {\overline X}_3 - {\overline X}_0  \right) 
- p_1^2 A(p_1) X_3
- B(p_1) X_2 - B(p_2){\overline X}_1 \,; 
\nonumber\\ 
C_3  &=& A(p_2){\overline X}_2 - A(p_1) X_2 - B(p_1) X_3 + B(p_2) {\overline X}_3 \,.
\label{Ci} 
\eea
On the other hand, contracting Eq.~(\ref{Li}) with $p_3^{\mu}$, we have 
\be
p_3^{\mu}\Gamma_{\mu}(p_1,p_2,p_3) = 
(p_2^2 - p_1^2) L_3 \mathbb{I} 
+[(p_2^2 - p_1^2) L_2 - L_1]\sla{p_1}
-[(p_2^2 - p_1^2) L_2 + L_1] \sla{p_2}
- 2 L_4 \tilde\sigma_{\mu\nu}p_1^{\mu} p_2^{\nu} \,.
\label{VLi}
\ee
Equating the right-hand sides of Eq.~(\ref{VCi}) and Eq.~(\ref{VLi}), we can express the 
$L_i$'s in terms of the functions $A$, $B$ and $X_i$'s. Specifically, 
\bea
L_1 &=& \frac{F(p_3)}{2} \bigg\{
A(p_1)[X_0 + (p_1^2- p_1\!\cdot\!p_2)X_3] 
+ A(p_2)[{\overline X}_0 +(p_2^2- p_1\!\cdot\!p_2){\overline X}_3]\bigg\} 
\nonumber\\
&+&
\frac{F(p_3)}{2} \bigg\{ B(p_1)(X_1+X_2) + B(p_2)({\overline X}_1+{\overline X}_2)\bigg\} \,;
\nonumber\\
L_2 &=& \frac{F(p_3)}{2(p_2^2 - p_1^2)} \bigg\{
A(p_1)[(p_1^2 + p_1\!\cdot\!p_2)X_3 -X_0] 
- A(p_2)[(p_2^2+p_1\!\cdot\!p_2){\overline X}_3 -{\overline X}_0]\bigg\}
\nonumber\\
&+&
\frac{F(p_3)}{2(p_2^2 - p_1^2)} \bigg\{ B(p_1)(X_2-X_1) + B(p_2)({\overline X}_1-{\overline X}_2)\bigg\} \,;
\nonumber\\
L_3 &=& - \frac{F(p_3)}{p_2^2 - p_1^2}
\bigg\{  
A(p_1) \left( p_1^2 X_1 + p_1\!\cdot\!p_2 X_2 \right)
- A(p_2) \left( p_2^2 {\overline X}_1 + p_1\!\cdot\!p_2 {\overline X}_2\right)
+ B(p_1)X_0 - B(p_2){\overline X}_0\bigg\} \,;
\nonumber\\
L_4 &=&\frac{F(p_3)}{2} \bigg\{ 
A(p_1) X_2 - A(p_2) {\overline X}_2 + B(p_1) X_3 - B(p_2){\overline X}_3 
\bigg\} \,.
\label{expLi}
\eea
The  standard approximation in the literature 
is to use the tree-level value 
of $H(p_1,p_2,p_3)$, which is equivalent to setting $X_0 = {\overline X}_0=1$ and $X_i = {\overline X}_i=0$, 
for $i \leq 1$, 
in the above equation. In addition, the effects 
of the ghost-dressing 
are also neglected, by imposing $F(p_3) =1$.
In this limit, we obtain the following expressions  
\bea
L_1 &=& \frac{A(p_1)+A(p_2)}{2} \,;
\nonumber\\
L_2 &=& \frac{A(p_1)- A(p_2)}{2(p_1^2 - p_2^2)} \,;
\nonumber\\
L_3 &=& \frac{B(p_1)- B(p_2)}{p_1^2 - p_2^2} \,; 
\nonumber\\
L_4 &=& 0 \,;
\eea
which leads to the so-called Ball-Chiu (BC) vertex~\cite{Ball:1980ay}
\bea 
\Gamma^{\mu}_{\rm BC}(p_1,p_2,p_3)&=&\frac{A(p_1)+A(p_2)}{2}\gamma^{\mu} \nonumber \\
&+&\frac{(p_1-p_2)^{\mu}}{p_1^2-p_2^2}\left\{\left[A(p_1)-A(p_2)\right] 
\frac{\sla{p_1}-\sla{p_2}}{2}
+\left[B(p_1)-B(p_2)\right] 
\right\} \,.
\label{bcvertex}
\eea
%


We will next insert into Eq.~(\ref{rsenergy1})
the general quark-gluon vertex of Eq.~(\ref{Li}) with the 
expressions for the form factors $L_i$ given in Eq.~(\ref{expLi}). Defining \mbox{$p_1=-p$}, \mbox{$p_2=k$}, and 
\mbox{$p_3=q$} and taking appropriate traces,  
it is straightforward to derive the 
following expressions for the integral equations satisfied by $A(p^2)$ and $B(p^2)$ 
\bea
 p^2 A(p^2)  &=& p^2 -  Z_c^{-1}C_{r}g^2 \left\{ \int_{k} \frac{L_1 (2 p_{\mu} p_{\nu} - k \!\cdot\!p \,g_{\mu\nu})
+2 (k^2+p^2) L_2 p_{\mu} p_{\nu}}{A^2(k^2)k^2-B^2(k^2)}
\,\Delta^{\mu\nu}(q) A(k^2)\right.
\nonumber\\
&+& \left. \int_{k} 
\frac{ p \!\cdot\!(k+p)L_4 g_{\mu\nu} -2 (L_3 + L_4) p_{\mu} p_{\nu}}{A^2(k^2)k^2-B^2(k^2)}
\,\Delta^{\mu\nu}(q) B(k^2)\right\} \,,
\eea

\bea
 B(p^2)  &=&  Z_c^{-1}C_{r}g^2 \left\{\int_{k} \frac{2 (L_4-L_3) p_{\mu} p_{\nu} - k \!\cdot\!(k+p) L_4 g_{\mu\nu}}
{A^2(k^2)k^2-B^2(k^2)}\,\Delta^{\mu\nu}(q) A(k^2)\right.
\nonumber\\
&+& \left.\int_{k} 
\frac{L_1 g_{\mu\nu} + 4 L_2 p_{\mu} p_{\nu}}{A^2(k^2)k^2-B^2(k^2)}
\,\Delta^{\mu\nu}(q)B(k^2)\right\} \,.
\eea

Using that, for the gluon propagator in the Landau gauge, \mbox{$\Delta^{\mu}_{\mu}(q) = -i 3 \Delta(q)$}, and
\mbox{$p_{\mu} p_{\nu} \Delta^{\mu\nu}(q) = -i h(p,k)\Delta(q)$}, 
where
\be
h(p,k) \equiv \frac{\left[k^2p^2-(k\!\cdot\!p)^2\right]}{q^2} \,,
\label{hf}
\ee
we can cast the gap equation in the form 
\bea
p^2 A(p)  &=& p^2 + i Z_c^{-1}C_{r}g^2 \left\{ \int_{k} \frac{2[(L_1 +(k^2+p^2) L_2]h(p,k) - 3 (k \!\cdot\!p) L_1}
{A^2(k^2)k^2-B^2(k^2)}
\,\Delta(q) A(k^2)\right.
\nonumber\\
&+& \left. \int_{k} 
\frac{ 3p \!\cdot\!(k+p)L_4 -2 (L_3 + L_4)h(p,k)}{A^2(k^2)k^2-B^2(k^2)}
\,\Delta(q) B(k^2)\right\} \,,
\label{gpa}
\eea

\bea
 B(p)  &=& -i Z_c^{-1}C_{r}g^2 \left\{\int_{k} \frac{2 (L_4-L_3)h(p,k) - 3 k \!\cdot\!(k+p) L_4 ]}
{A^2(k^2)k^2-B^2(k^2)}\,\Delta(q) A(k^2)\right.
\nonumber\\
&+& \left.\int_{k} 
\frac{3L_1  + 4 h(p,k)L_2}{A^2(k^2)k^2-B^2(k^2)} 
\,\Delta(q)B(k^2)\right\} \,.
\label{gpb}
\eea 

\subsection{The ``one-loop dressed'' approximation for $X_0$}
\label{ss2}

In order to evaluate Eqs.~(\ref{gpa}) and (\ref{gpb}) further, it is necessary 
to determine the behavior of the  form factors $X_i$ entering 
into the definition of the $L_i$ through the Eq.~(\ref{expLi}).

From Eq.~(\ref{Xi}), one can  projected out the form factors $X_i$ 
in the following way
\bea
X_0 &=& \frac{\rm Tr\{H\}}{4} \,;
\nonumber\\
X_1 &=& \frac{ p_2^2 {\rm Tr}\{\sla{p_1}H\} - (p_1\!\cdot\!p_2) {\rm Tr}\{\sla{p_2}H\}}
{4 \left[p_1^2 p_2^2 - (p_1\!\cdot\!p_2)^2\right]}\,;
\nonumber\\
X_2 &=& \frac{ p_1^2 {\rm Tr}\{\sla{p_2}H\} - (p_1\!\cdot\!p_2) {\rm Tr}\{\sla{p_1}H\}}
{4 \left[p_1^2 p_2^2 - (p_1\!\cdot\!p_2)^2\right]}\,;
\nonumber\\
X_3 &=& \frac{ {\rm Tr}\{\tilde\sigma_{\alpha\rho}\, p_1^{\alpha} p_2^{\rho}\, H\}}
{4 \left[(p_1\!\cdot\!p_2)^2 -p_1^2 p_2^2\right]}\,.
\label{invX}
\eea
It is clear from the diagrammatic representation given in 
Fig.~\ref{fig_sk} that $H(p_1,p_2,p_3)$, and its form factors given in  Eq.~(\ref{invX}),
depend 
among other things, on the fully dressed quark propagator. Therefore, 
the treatment of the full gap equation given in  Eqs.~(\ref{gpa}) and (\ref{gpb}) 
boils down to a complicated system of several coupled integral equations.  
In order to make the problem at hand technically more tractable, 
we will only retain one of the form factors given in  Eq.~(\ref{invX}), 
and study it in an approximate kinematic configuration, which simplifies the 
resulting structures considerably. Specifically, we will
only consider the form factor $X_0$, and assume that 
$p_1 = p_2 \equiv p$, and  $p= - p_3/2$,  
so that $X_0 = {\overline X}_0$.

\begin{center}
\begin{figure}[t]
\includegraphics[scale=0.5]{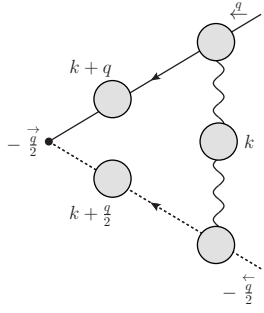}
\caption{Diagrammatic representation of the quark-ghost scattering kernel, $H(-q/2,-q/2,q)$, at one-loop.}
\label{fig_sk_loop}
\end{figure}
\end{center}

Note that the momentum $p_3$ coincides with the momentum of the 
gluon in the gap equation; therefore we will denote $p_3 = q$, and $p= - q/2$.   
To obtain a non-perturbative estimate for $X_0$, we will study the 
``one-loop dressed'' contribution given by the diagram of Fig.~\ref{fig_sk_loop}.

\vspace{-1.5cm}

Denoting  the full gluon-ghost vertex by $G_{\mu}^{ab}=\delta^{ab} G_{\mu}$,                        
the expression for $X_0^{[1]}$ reads 
\be
X_0^{[1]}(p,p,q) = 1 - i\left(\frac{1}{4}\right) \frac{C_Ag^2}{2} \int_k \Delta^{\mu\nu}(k)D(k-p)\, G_{\nu}\, 
{\rm Tr}\{ S(k+q) \Gamma_{\mu}\} \,.
\label{X0LD}
\ee

To evaluate this further, we will use the following approximations:
(i) $G_{\nu}$ will be replaced by its tree-level value, $G_{\nu} = (k-p)_{\nu}$.
Note that, since  the full $\Delta^{\mu\nu}(k)$ is transverse (Landau gauge), 
only the  $p_{\mu}$ part survives. (ii) For the vertex $\Gamma_{\mu}$ we will use a slightly modified Ansatz than 
that of Eq.~(\ref{bcvertex})~\cite{Atkinson:1988mw,Krein:1988sb,Aguilar:2005sb}, in order to 
reduce the algebraic complexity of the resulting equation.

Specifically, we will use as our starting point the Ansatz
\be
\Gamma_{\mu}(q,-k-q,k) = \frac{1}{2}\left( [A(k+q)+A(q)]\gamma_{\mu} + 
\frac{k_{\mu}}{k^2}[ A(k+q)-A(q)](2 \sla{q}+\sla{k}) \right) \,, 
\label{Atk}
\ee
which satisfies the STI of Eq.~(\ref{STI}) in the chiral limit, $B=0$, and with 
the ghost sector switched off ($F=H=1$). 
  Since the second term on the rhs of Eq.~(\ref{Atk})
is proportional to $k_{\mu}$, 
it vanishes when inserted into~(\ref{X0LD}), again due to the transversality of $\Delta^{\mu\nu}(k)$,
a fact that simplifies considerably the resulting expressions. 

According to the procedure discussed in the next section [see Eq.~(\ref{BCCPimp})], we will improve 
the Ansatz of Eq.~(\ref{Atk}) by multiplying it by $F(k)$, but keeping $H=1$.    
Under these approximations, and after 
setting $p= - q/2$, Eq.~(\ref{X0LD}) becomes (in Euclidean space) 
\be
X_0^{[1]}(q)= 1+ \frac{C_Ag^2}{8} \int_k \left[q^2-\frac{(q\!\cdot\!k)^2}{k^2}\right]\Delta(k)
D\left(k+\frac{q}{2}\right) F(k) \, 
\frac{A(k+q) [A(k+q)+A(q)]}{A^2(k+q) (k+q)^2 + B^2(k+q)} \,.
\label{X0LD1}
\ee
We will finally approximate  $A(k+q)=A(q)=1$, and $B(k+q)=0$, then we obtain 
\be
X_0^{[1]}(q) = 1 + \frac{1}{4} C_A g^2q^2 \int_k [1-f(k,q)]\Delta (k) F(k) \frac{F(k+q)}{(k+q)^4} \,,
\label{sk2}
\ee
where 
\be
f(k,q) \equiv  \frac{(k \cdot q)^2}{k^2 q^2}\,.
\ee
After carrying out the integral on the rhs of 
Eq.(\ref{sk2}) one obtains an approximate expression for $X_0^{[1]}(q)$ in terms of 
$\Delta (k)$ and $F(k)$; the result will be reported in section \ref{numan}. 
Note that if we had  
multiplied  Eq.~(\ref{Atk}) not only by $F(q)$ but also by $X_0^{[1]}(q)$, as is done in Eq.~(\ref{BCCPimp}), 
then, instead of simply computing the  integral of Eq.~(\ref{sk2}),  
one would have to deal with the more difficult task 
of solving an integral equation for the unknown $X_0^{[1]}(q)$.

\subsection{The gap equation with ghost-improved quark-gluon vertices}
\label{ss3}

There are two basic forms of the ``abelianized'' quark-gluon vertex usually employed 
in the literature: (i) the BC vertex,  denoted by $\Gamma^{\mu}_{\rm BC}$, whose closed 
form is given in Eq.~(\ref{bcvertex}), and  
(ii) the so-called 
Curtis and Pennington (CP) vertex~\cite{Curtis:1990zs}, to be denoted by $\Gamma^{\mu}_{\rm CP}$.
These two vertices differ by a transverse (automatically conserved term), namely 
\bea
\Gamma^{\mu}_{\rm CP}(p_1,p_2,p_3) = \Gamma^{\mu}_{\rm BC}(p_1,p_2,p_3) 
+ \Gamma^{\mu}_{\rm T}(p_1,p_2,p_3)\,, 
\label{full_cp1}
\eea
with 
\bea
\Gamma^{\mu}_{\rm T}(p_1,p_2,p_3) = \frac{\gamma^{\mu}(p_2^2-p_1^2) -(p_1-p_2)^{\mu}(\sla{p_1}+\sla{p_2})}{2d(p_1,p_2)} 
\left[A(p_2)-A(p_1)\right] \,,
\label{cp_tran}
\eea
where 
\bea
d(p_1,p_2) = \frac{1}{p_1^2+p_2^2}\left\{(p_2^2-p_1^2)^2 + \left[\frac{B^2(p_2)}{A^2(p_2)}+\frac{B^2(p_1)}{A^2(p_1)}\right]^2\right\}\,.
\eea
Evidently, under the approximations employed,  
the ghost effects due to $F(p_3)$ and $X_0^{[1]}(p_3)$ 
may be incorporated into these two vertices through simple multiplication of their 
form factors by  $F(p_3)X_0^{[1]}(p_3) $.
Denoting their ``ghost-improved'' versions by 
$\overline\Gamma^{\mu}_{\rm CP}(p_1,p_2,p_3)$ and $\overline\Gamma^{\mu}_{\rm CP}(p_1,p_2,p_3)$, respectively,
we have 
\bea
\overline\Gamma^{\mu}_{\rm BC}(p_1,p_2,p_3) &=& F(p_3) X_0^{[1]}(p_3) \Gamma^{\mu}_{\rm BC}(p_1,p_2,p_3)\,,
\nonumber\\
\overline\Gamma^{\mu}_{\rm CP}(p_1,p_2,p_3) &=&  F(p_3) X_0^{[1]}(p_3)\Gamma^{\mu}_{\rm CP}(p_1,p_2,p_3)\,. 
\label{BCCPimp}
\eea
Notice that, one recovers the vertex used in Ref.~\cite{Fischer:2003rp} by setting $X_0^{[1]}(q)=1$ in the
above equations.
 
Substituting the form factors of 
$\overline\Gamma^{\mu}_{\rm BC}(p_1,p_2,p_3)$ into Eqs.~(\ref{gpa}) and (\ref{gpb}), we 
arrive at the following coupled system for $A(p^2)$ and $B(p^2)$ 
\bea
A(p^2)&=& 1 + C_{r}g^2  Z_c^{-1}\,\int_{k}\,
\frac{{\cal K}_0(p-k)}{A^2(k^2)k^2+B^2(k^2)}{\cal K}_A^{\rm BC}(k,p)\,, 
\label{dirac}\\ 
B(p^2) &=& C_{r}g^2  Z_c^{-1}\int_{k}\,\frac{{\cal K}_0(p-k)}{A^2(k^2)k^2+B^2(k^2)} {\cal K}_B^{\rm BC} (k,p)\,,
\label{scalar}
\eea
where the kernel ${\cal K}_0(q)$ corresponds to the part 
that is not altered by the tensorial structure of the quark-gluon vertex, namely
\bea
{\cal K}_0(q) =\Delta(q)F(q)X_0^{[1]}(q) \,, 
\label{ker1}
\eea
while the parts that are affected, ${\cal K}_A^{\rm BC}(k,p)$ and ${\cal K}_B^{\rm BC}(k,p)$, 
are given by~\cite{Hawes:1993ef}
\bea
{\cal K}_A^{\rm BC} (k,p) &=& \frac{A(k^2)}{2p^2}[A(k^2)+A(p^2)]
\left[3p\!\cdot\!k -2h(p,k) \right] -2B(k^2)\Delta B(k^2,p^2) \frac{h(p,k)}{p^2}   \nonumber \\ 
&&   -A(k^2)\Delta A(k^2,p^2)\left[k^2 -\frac{(k\!\cdot\!p)^2 }{p^2} +2\frac{k\!\cdot\!p }{p^2}h(p,k)\right]  \,,
\\
{\cal K}_B^{\rm BC} (k,p) &=& \frac{3}{2} B(k^2)[A(k^2)+A(p^2)] + 
2\left[B(k^2)\Delta A(k^2,p^2) -A(k^2)\Delta B(k^2,p^2) \right]h(p,k) \,, \nonumber
\label{kernels}
\eea
where $h(p,k)$ is given by Eq.~(\ref{hf}) and
\be
\Delta A(k^2,p^2) \equiv \frac{A(k^2)-A(p^2)}{ k^2-p^2} \,,\,\qquad
\Delta B(k^2,p^2) \equiv \frac{B(k^2)-B(p^2)}{ k^2-p^2} \,.
\label{delta}
\ee

Similarly, the effect of the vertex $\overline\Gamma^{\mu}_{\rm CP}(p_1,p_2,p_3)$ is 
to replace the kernels   
${\cal K}_A^{\rm BC}(k,p)$ and ${\cal K}_B^{\rm BC}(k,p)$  
appearing in Eq.~(\ref{dirac}) and (\ref{scalar}) by ${\cal K}_A^{\rm CP}(k,p)$ and ${\cal K}_B^{\rm CP}(k,p)$, respectively, where
\bea
{\cal K}_A^{CP}(k,p) &=&  {\cal K}_A^{BC}(k,p) + \frac{3k\cdot p}{2p^2}A(k^2)\Delta A(k^2,p^2)\frac{(k^2-p^2)^2}{d(k,p)} \,, \nonumber \\ 
{\cal K}_B^{CP}(k,p) &=&  {\cal K}_B^{BC}(k,p) + \frac{3}{2}B(k^2)\Delta A(k^2,p^2)\frac{(k^2-p^2)^2}{d(k,p)} \,.
\label{kcp}
\eea 

Note that the above equations 
are written in the Euclidean space. Specifically, the Wick rotation
was performed  by setting \mbox{$p^2= -p^2_{\da}$},
\mbox{$\de(p^2_{\da})= -  {\Delta}(-p^2_{\da})$}, \mbox{$A(-p^2_{\da})=A_{\da}(-p^2_{\da})$}, 
and \mbox{$B(-p^2_{\da})=B_{\da}(-p^2_{\da})$}. Therefore,
 $\Delta A(k^2,p^2)$, $\Delta B(k^2,p^2)$,  and $h(p,k)$  change the sign under Wick rotation. For the  integration  measure, we used 
\mbox{$\int_k=i\int_{k_\mathrm{E}}$};   as a last step, we have suppressed the subscript ``E'' everywhere.

\subsection{Asymptotic behavior and renormalization group properties}
\label{ss4}

The study  of the ultraviolet (UV) behavior of the dynamical quark mass, ${\mathcal{M}}(p^2)$, predicted
by the coupled system formed by Eqs.~(\ref{dirac}) and (\ref{scalar}), reveals the need 
of one final adjustment. 
Specifically, as is well known, 
the correct UV behavior for ${\mathcal{M}}(p^2)$ is given by~\cite{Miransky:1981rt, Politzer:1976tv} 
\be
{\mathcal M}(x) = \frac{c}{x}\left[\ln\left(\frac{x}{\Lambda^2}\right)\right]^{\gamma_f-1} \,,
\label{asy}
\ee
where $c$ is a 
$\mu$-independent constant, related to the chiral condensate 
\mbox{$\left\langle \bar{q}q\right\rangle_{\mu}$} by~\cite{Hawes:1993ef} 
\be
c= -\frac{4\pi^2\gamma_f}{3} \left\langle \bar{q}q\right\rangle_{\mu}
\left[\ln\left(\frac{\mu^2}
{\Lambda^2}\right)\right]^{-\gamma_f} \,.
\label{const}
\ee
and the mass anomalous dimension is given by
\mbox{ $\gamma_f = 9 C_{\rm r}/(11C_{\rm A} -3/2 C_{\rm r}n_f)$}. For the fundamental
representation, we have $C_{\rm r}=4/3$ and then \mbox{ $\gamma_f = 12/(11C_{\rm A} -2n_f)$}. 
Equivalently, we can rewrite it in terms of the first coefficient $b = (11C_{\rm A} -2n_f)/(48\pi^2)$ of the QCD $\beta$ function,
where \mbox{$\gamma_f = 3C_{\rm F}/(16\pi^2 b)$}.  

On the other hand, after setting $Z_1=1$ and $X_0^{[1]}=1$, neglecting $\Delta A(k^2,p^2)$ and $\Delta B(k^2,p^2)$, 
and setting \mbox{$A(p^2)=A(k^2) \to 1$}, 
the asymptotic equation that the system (\ref{dirac})-(\ref{scalar}) yields for 
${\mathcal{M}}(p^2)$ is given, after the standard angular approximation, by 
\begin{equation}
{\mathcal M}(p^2) =\frac{3 C_{\rm r}g^2}{16\pi^2}
\left[\Delta(p^2)F(p^2)\int_0^{p^2}\!\!\! dk^2\, {\mathcal M}(k^2)  + 
\int_{p^2}^{\infty}\!\!\! dk^2\, \Delta(k^2)F(k^2) {\mathcal M}(k^2)
\right]\,. 
\label{asy_eq}
\end{equation}

In order to we verify whether indeed ${\mathcal{M}}(p^2)$ of Eq.~(\ref{asy})
satisfies Eq.~(\ref{asy_eq}), we substitute ${\mathcal{M}}(p^2)$ into the rhs
of Eq.~(\ref{asy_eq}), together with the one-loop gluon propagator
\be
\Delta^{-1}(p^2)=  p^2\left[1+  \gamma_1g^2\ln\left(\frac{p^2}{\mu^2}\right)\right]\,,
\label{1loop_g}
\ee  
where $\gamma_1= (\frac{13}{3}C_{\rm A} - C_{\rm{r}}n_f)/32\pi^2$, and the one-loop ghost
dressing function 
\be 
F^{-1}(p^2)= \left[1+\frac{9}{4}\frac{C_{\rm A}g^2}{48\pi^2}\ln\left(\frac{p^2}{\mu^2}\right) \right] \,.
\label{fasg}
\ee 

After performing the above substitutions, it is straightforward 
to see that for asymptotic large values of $p^2$, the dominant contribution comes from the 
first integral, and that the solution 
of Eq.~(\ref{asy_eq}) is indeed of the form given in~(\ref{asy}), but with one important difference:
the anomalous dimension assumes the value $\gamma_f= 36 C_{\rm r}/(35C_{\rm A} -6C_{\rm r}n_f)$, instead of the 
correct value given above.
Note that the difference is due to the non-Abelian contributions (gluons); 
setting $C_A =0$, the two values coincide.  

The reason for this discrepancy can be traced back to a typical ambiguity, intrinsic to the 
gauge-technique. Specifically, the standard procedure of constructing a vertex 
based on the requirement that it should satisfy the correct STI, 
leaves  the transverse  (i.e.    automatically  conserved)  part   of  the  vertex
undetermined~\cite{Salam:1963sa, Kizilersu:2009kg}.  While, in the presence  of mass gaps, 
the infrared dynamics appear to be largely unaffected,
this ambiguity is
known to modify the 
ultraviolet  properties of the SD equations~\cite{Salam:1963sa}.   
Essentially, failing  to provide  the correct
transverse  part  leads  to  the mishandling of  overlapping  divergences,
which,  in turn, compromises  the multiplicative  renormalizability of
the resulting SD equations.
The construction of the appropriate transverse part is technically complicated even for QED~\cite{Curtis:1990zs},
and its generalization to a non-abelian context is still pending. 
Given the expectation that the restoration of the transverse part should not 
affect the infrared dynamics, 
the usual short-cut employed in the literature 
is to account approximately for the missing pieces by modifying (by hand) the SDE in question.

Specifically, for the case at hand,
the correct asymptotic behavior for ${\mathcal{M}}(p^2)$ 
may be restored by carrying out in Eqs.~(\ref{dirac}) and (\ref{scalar})
the replacement 
\be
 Z_c^{-1} {\cal K}_{A,B} (k,p) \to  {\cal K}_{A,B} (k,p) F(p^2)\,.
\label{rgi_improve}
\ee 
given that, for large $p^2$, the perturbative (one-loop) expression  for $F(p^2)$ is given by Eq.(\ref{fasg}).

It is straightforward to verify that, with this modification, Eq.~(\ref{asy_eq})
yields the correct value for $\gamma_f$. Note that even though, strictly speaking, one needs 
to supply only the asymptotic form given in~(\ref{fasg}), it is natural to assume 
that the non-perturbative completion of this formula will be given by Eq.~(\ref{ghost}), namely 
the full  $F(p^2)$. 

At first sight it would seem that the modification introduced in Eq.~(\ref{rgi_improve})
amounts to the counter-intuitive replacement  
$ Z_c^{-1} \to  F(p^2)$, i.e. trading a (cutoff-dependent) constant for a 
($\mu$-dependent) function of $p^2$~\cite{Fischer:2003rp}. 
Even though 
from the operational point of view this is indeed what happens, the idea behind is slightly more 
subtle. A more intuitive way to interpret Eq.~(\ref{rgi_improve}) 
is that the corresponding kernels are modified 
due to the presence of the (unknown)  transverse parts, 
whose additional contributions must be such that, when properly renormalized,  
will effectively amount to the replacement given in Eq.~(\ref{rgi_improve}). 
Needless to say, it would be very important to determine the precise mechanism that 
leads to these modifications, but this is at present beyond our powers.

It is instructive to understand from a slightly different 
(but ultimately equivalent) point of view why the (minimal)  
factor that must be supplied to Eq.~(\ref{asy_eq})
is indeed $F(p^2)$.
The argument relies on the (expected) RG-invariance of 
the mass function ${\mathcal{M}}(p^2)$. 
Specifically, let us for the moment assume that 
the underlying theory is QED, and set $F=1$ into Eq.~(\ref{asy_eq}). 
Then, ${\mathcal{M}}(p^2)$ is RG-invariant, because, 
in QED, due to the abelian Ward identities, the product $g^2\Delta$ 
appearing on the rhs of  Eq.~(\ref{asy_eq}) 
is RG-invariant. In the case of QCD this is no longer true in general; 
instead, in the $R_{\xi}$-type of gauges the corresponding RG-invariant combination  
is given by $g^2\Delta F^2$~\cite{Fischer:2003rp, Aguilar:2009nf}. 
Therefore, the simplest way to convert the product $g^2\Delta F$
into a RG-invariant combination is to 
multiply it by $F$, which restores the 
RG-invariance of ${\mathcal{M}}(p^2)$.

\section{Numerical analysis}
\label{numan}

In this section  we will first review some of the recent lattice data for
the gluon propagator $\Delta(q)$ and the ghost dressing function $F(q)$.
Then, we  will substitute them into Eq.~(\ref{sk2}), in order to 
to  obtain an estimate for $X_0^{[1]}(q)$. 
With all necessary ingredients \mbox{[$\Delta(q)$, $F(q)$, and $X_0^{[1]}(q)$]}   available, we proceed to  
solve numerically the coupled system formed by 
the integral equations (\ref{dirac}) and (\ref{scalar}) when
the non-abelian modifications of the BC and CP vertices are implemented. 
We will solve the corresponding equations both for fermions 
in the fundamental (quarks)  and in the adjoint representations; for the former case 
we will use the numerical results obtained to compute the 
pion decay constant and the quark condensate.

\subsection{Nonperturbative ingredients: gluon and ghost propagators from lattice and
the form factor $X_0^{[1]}(q)$}

\begin{figure}[!t]
\begin{minipage}[b]{0.45\linewidth}
\centering
\includegraphics[scale=0.4]{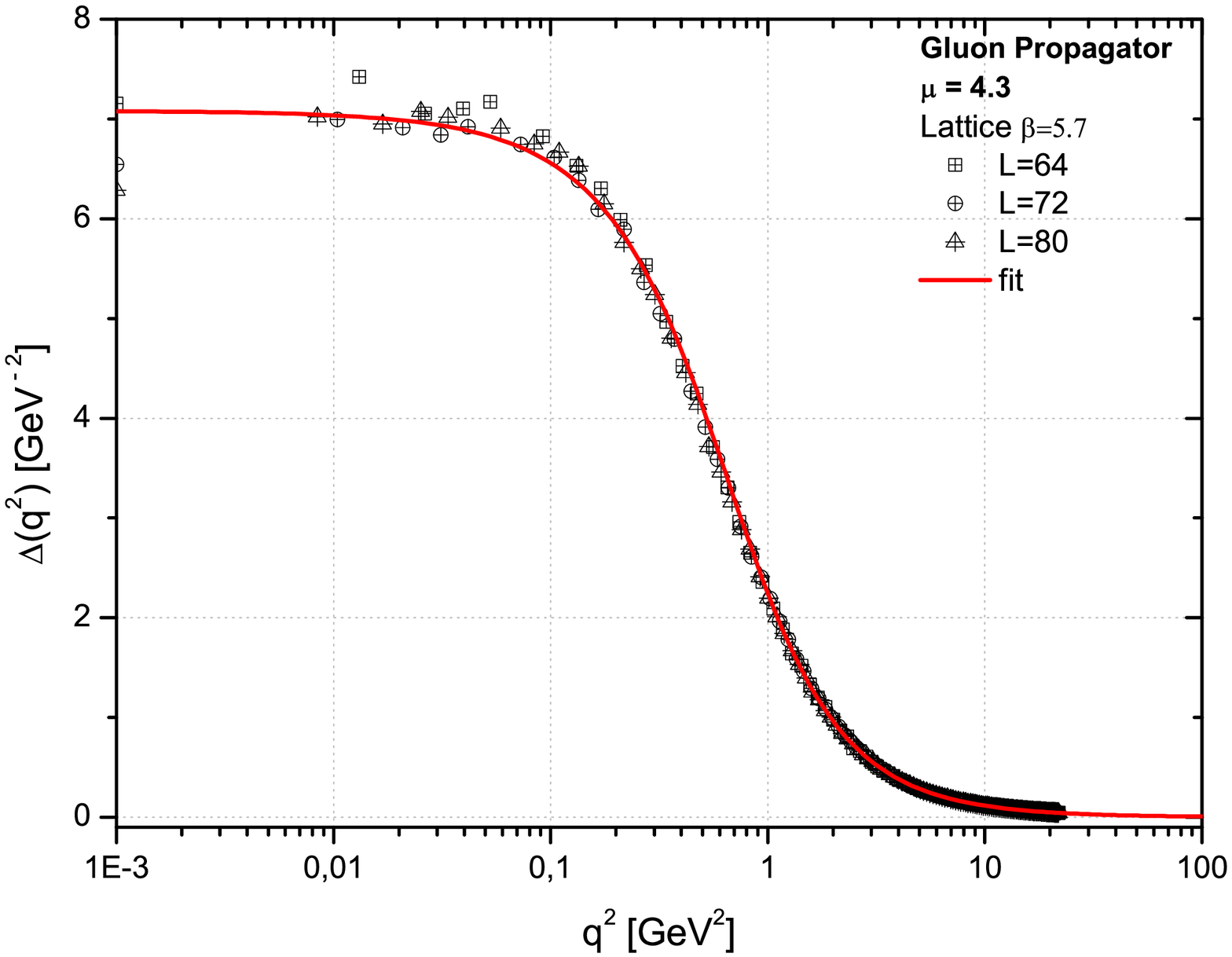}
\end{minipage}
\hspace{0.5cm}
\begin{minipage}[b]{0.50\linewidth}
\includegraphics[scale=0.4]{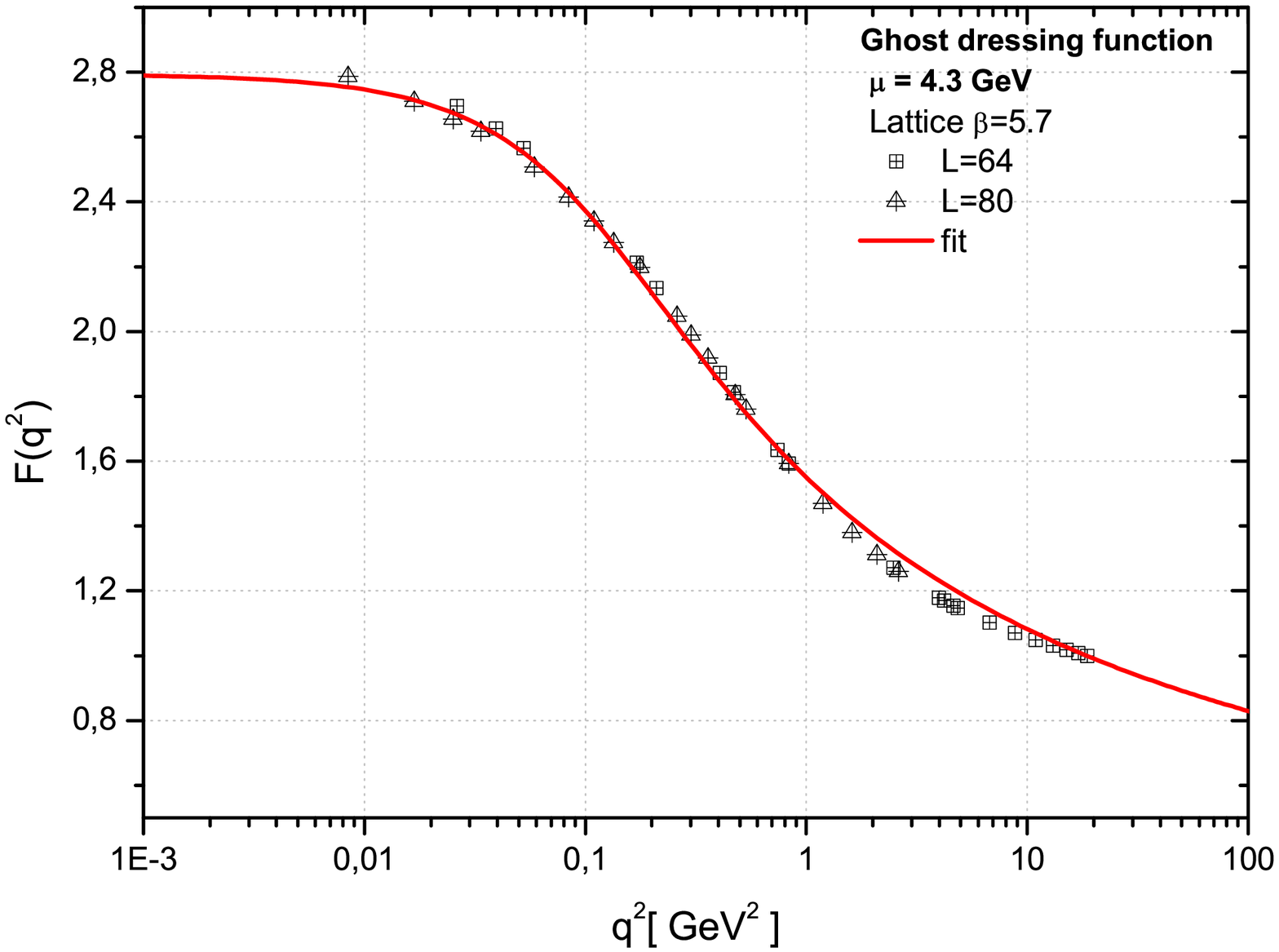}
\end{minipage}
\vspace{-1.0cm}
\caption{Lattice results for the gluon propagator, $\Delta(q)$, and ghost dressing, $F(q)$,
obtained in  Ref.~\cite{Bogolubsky:2007ud} and renormalized at $\mu=4.3$ GeV. 
{\it Left panel}: The continuous line represents the gluon lattice data
fitted by Eq.~(\ref{gluon}) when \mbox{$m= 520\,$ MeV},
\mbox{$g_1^2=5.68$}, \mbox{$\rho_1=8.55$} and, \mbox{$\rho_2=1.91$}.  
{\it Right panel}: The lattice data for $F(q)$ fitted by Eq.~(\ref{ghost}) using 
\mbox{$g_2^2 = 8.57$},
\mbox{$m = 520\,$ MeV}, \mbox{$\rho_3=0.25$} and, \mbox{$\rho_4=0.68$}.}
\label{fig1}
\end{figure}

In this subsection we will comment on the nonperturbative form 
of the three basic Green's functions
entering into the gap equation, namely $\Delta(q)$, $F(q)$, and $X_0^{[1]}(q)$.
 
We start by showing on the left panel of Fig.~\ref{fig1} the lattice data for gluon propagator obtained
in~\cite{Bogolubsky:2007ud}. The lattice data presented there 
correspond to a $SU(3)$ quenched lattice simulation, where $\Delta(q)$
is renormalized at $\mu=4.3$ GeV.  
In this plot, we clearly see the appearance of a plateau in the deep IR region.
The IR finiteness of the gluon propagator has been long associated with the dynamical generation 
of an effective gluon mass~\cite{Cornwall:1982zr,Aguilar:2006gr, Aguilar:2008xm}. 
In fact, the above set of lattice data can be accurately fitted in 
terms of the following physically motivated expression~\cite{Aguilar:2010gm}
\be
\Delta^{-1}(q^2)= m^2(q^2) + q^2\left[1+ \frac{13C_{\rm A}g_1^2}{96\pi^2} 
\ln\left(\frac{q^2 +\rho_1\,m^2(q^2)}{\mu^2}\right)\right]\,,
\label{gluon}
\ee  
with $m^2(q^2)$  given by
\be
m^2(q^2) = \frac{m^4}{q^2 + \rho_2 m^2} \,,
\label{dmass}
\ee
where the fitting parameters are 
\mbox{$m= 520$\,\mbox{MeV}}, \mbox{$g_1^2=5.68$}, \mbox{$\rho_1=8.55$} and, \mbox{$\rho_2=1.91$}.
The parameter $m$ acts as a physical  mass 
scale, whose function is  to regulate the perturbative RG logarithm; so, instead 
of diverging at the Landau pole, the 
logarithm saturates at a finite value~\cite{Aguilar:2010gm}. In addition, for large values of $q^2$, we
 recover the one-loop expression of the gluon propagator in the Landau gauge given by Eq.~(\ref{1loop_g}). 
Note that, contrary to conventional masses, dynamically generated masses  
display a  non-trivial dependence on the momentum transfer $q^2$~\cite{Cornwall:1982zr}. 
In particular, $m^2(q^2)$ assumes 
a non-zero value in the IR, and drops rapidly in the UV in 
a way consistent with the operator product expansion~\cite{Lavelle:1991ve,Aguilar:2007ie,Dudal:2008sp}.

On the right panel of Fig.~\ref{fig1}, we show the lattice results for $F(q)$ 
obtained from~\cite{Bogolubsky:2007ud}; evidently, $F(q)$
saturates in the deep IR at the constant value~\cite{Aguilar:2008xm,Boucaud:2008ji}. 
The ghost dressing function 
is also renormalized at $\mu=4.3$ GeV, and the  data can be accurately fitted by the expression 
\be
F^{-1}(q^2)= 1+ \frac{9}{4}\frac{C_{\rm A}g_2^2}{48\pi^2}
\ln\left(\frac{q^2 +\rho_3  m^2(q^2)}{\mu^2}\right)\,,
\label{ghost}
\ee
where the dynamical mass is given by Eq.~(\ref{dmass}) changing 
the parameter $\rho_2 \to \rho_4$. The fitting parameters are \mbox{$g_2^2 = 8.57$},
\mbox{$m = 520\,$ MeV}, \mbox{$\rho_3=0.25$} and, \mbox{$\rho_4=0.68$}.
%
\begin{center}
\begin{figure}[t]
\includegraphics[scale=0.4]{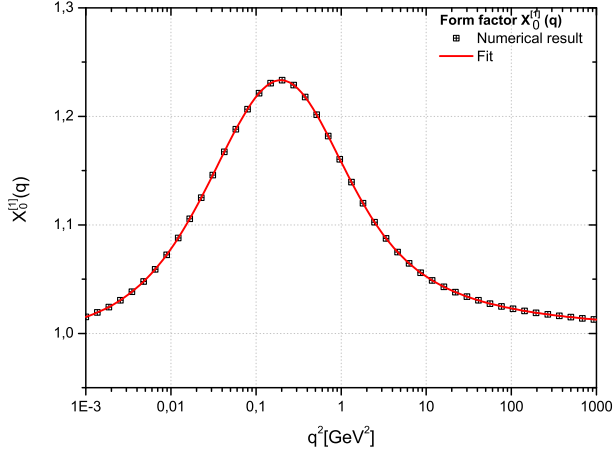}
\caption{Numerical result for the form factor $X_0^{[1]}(q)$ of  the  
 quark-ghost scattering kernel given by Eq.~(\ref{sk2}) when $\alpha(\mu^2)=0.295$.} 
\label{sk}
\end{figure}
\end{center}
\vspace{-1.5cm}

The last ingredient to be determined is the form factor $X_0^{[1]}(q)$. We proceed
substituting the fit for lattice data for $\Delta(q)$ and $F(q)$ presented in
Fig.~\ref{fig1} into Eq.~(\ref{sk2}). Then, for determining the integral  given by  Eq.~(\ref{sk2}), we should 
fix the value of $g^2(\mu^2)$.
We adopt the same procedure of Ref.~\cite{Aguilar:2010gm}, where
it was found that the perturbative tail of  effective coupling, determined from the lattice data,
is compatible with four-loop perturbative calculation 
at MOM scheme, presented in~\cite{Boucaud:2005rm}.  
More specifically, we use $\alpha(\mu^2)=g^2(\mu^2)/4\pi=0.295$.
The  numerical result for $X_0^{[1]}(q)$ is shown in the  
Fig.~\ref{sk}, and  it can be fitted by
\bea
a_2 + \frac{a_1 - a_2}{1+\left(q^2/q^2_0\right)^p} \,,
\eea  
where \mbox{$a_1 = 1.0$}, \mbox{$a_2=1.27$}, \mbox{$q^2_0=0.027 \,\mbox{GeV}^2$} and,  \mbox{$p=1.0$} when \mbox{$q^2\leq 0.202 \,\mbox{GeV}^2$}; while
for \mbox{$q^2 > 0.202 \,\mbox{GeV}^2$}, we choose \mbox{$a_1 = 1.40$}, \mbox{$a_2=1.0$}, \mbox{$q^2_0=0.45 \,\mbox{GeV}^2$} and,  \mbox{$p=0.63$}.
  
$X_0^{[1]}(q)$ shows a maximum  
located in the intermediate momentum region (around \mbox{$450$ MeV}),   
while in the UV and IR regions $X_0^{[1]}(q) \to 1$.
Although this peak is not very pronounced (notice the small scale in y-axis), we will see soon that 
it is  essential for providing to  
the kernel of the gap equation the enhancement required for the generation of 
phenomenologically compatible constituent quark masses.

\subsection{Chiral symmetry breaking in the fundamental representation 
and the pion decay constant}

\begin{figure}[!t]
\begin{minipage}[b]{0.45\linewidth}
\centering
\includegraphics[scale=0.4]{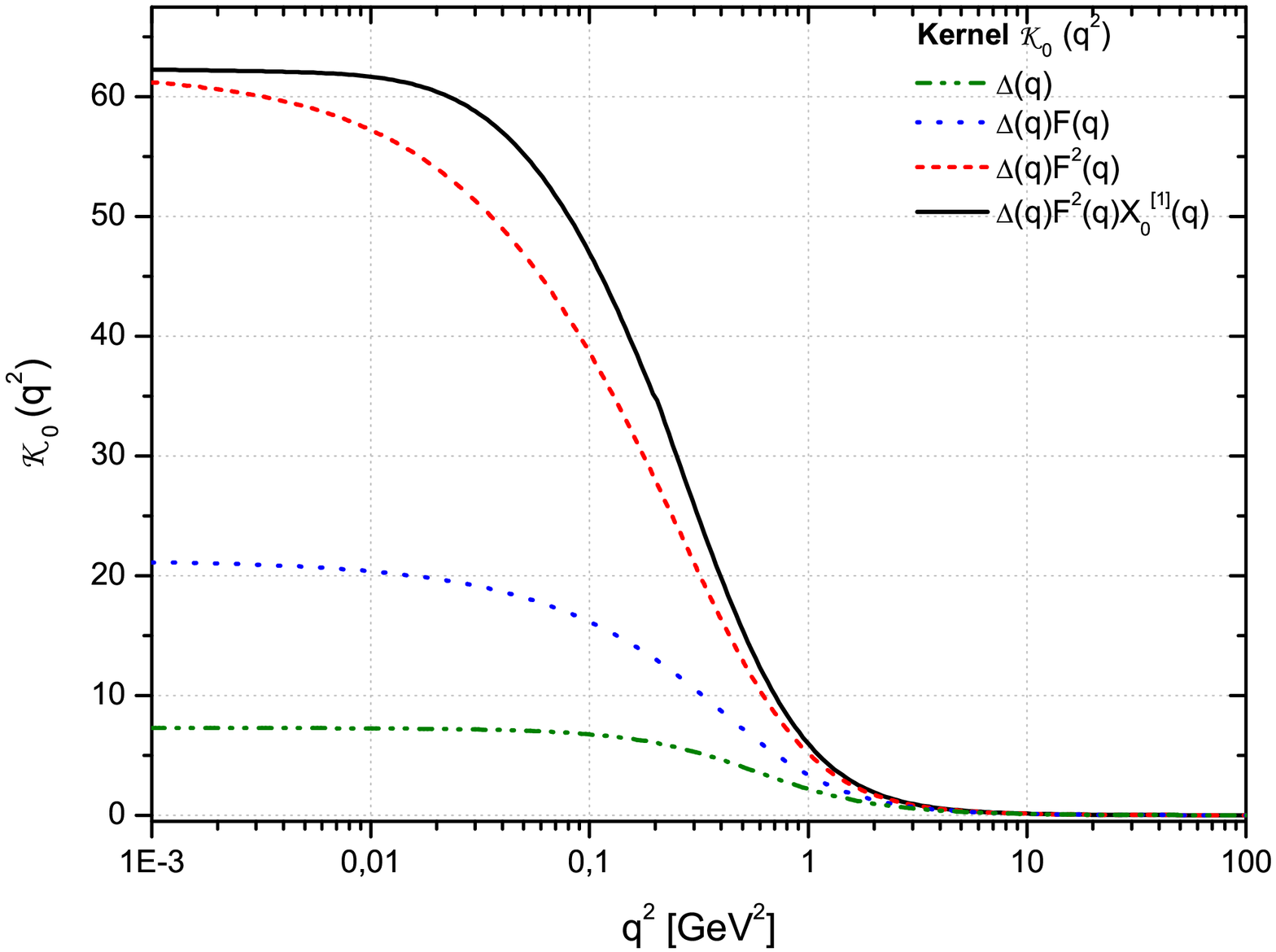}
\end{minipage}
\hspace{0.5cm}
\begin{minipage}[b]{0.50\linewidth}
\centering
\includegraphics[scale=0.4]{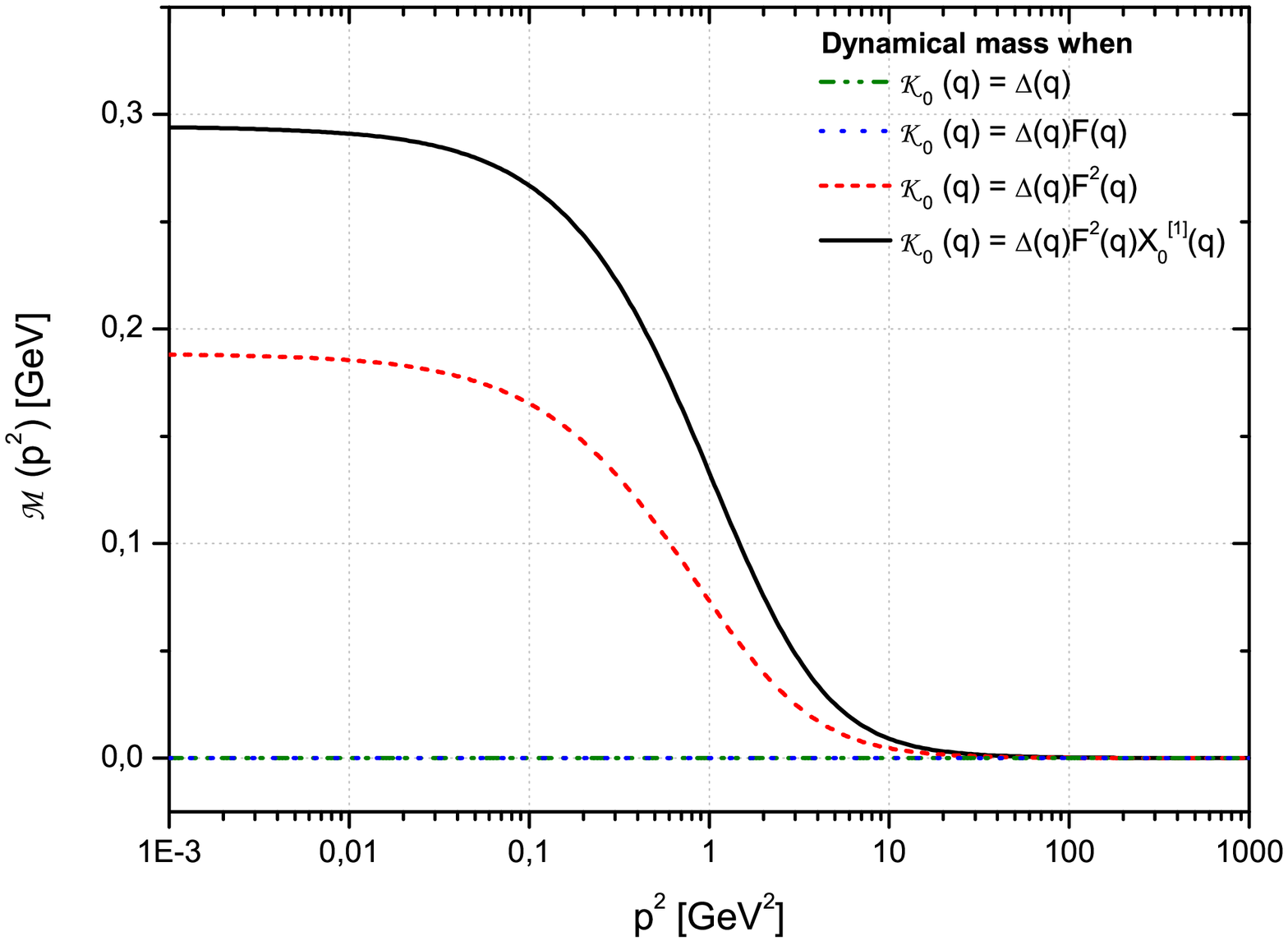}
\end{minipage}
\vspace{-1.0cm}
\caption{{\it Left panel}: The individual contribution of the 
ingredients composing $ {\cal K}_0(q)$. 
The green dotted-dashed line represents
the case where \mbox{${\cal K}_0(q)=\Delta(q)$}, the blue dotted line the 
 case where \mbox{${\cal K}_0(q)=\Delta(q)F(q)$}, in the red dashed curve 
\mbox{${\cal K}_0(q)=\Delta(q)F^2(q)$} and, finally the black 
continuous line represents the case where ${\cal K}_0(q)$ assumes 
the full form used in our calculation \mbox{${\cal K}_0(q)=\Delta(q)F^2(q)X_0^{[1]}(q)$}. 
{\it Right panel}: The corresponding 
dynamical quark mass generated when we use in Eqs.(\ref{dirac}) and (\ref{scalar}) 
the different forms of ${\cal K}_0(q)$ are shown in the left panel.}
\label{fig3}
\end{figure}

We now proceed to the solution of the coupled system of integral equations defined by the Eqs.(\ref{dirac}) and (\ref{scalar}).
Note that, in addition, and in accordance to the discussion given in subsection \ref{ss4}, we carry out the substitution 
\be
F(q) \to  F^2(q)\,,
\label{FtoF2}
\ee 
%
in the Eq.~(\ref{ker1}). In other words, the kernel ${\cal K}_0(q)$ appearing on the rhs of Eqs.(\ref{ker1}) and (\ref{scalar})
will assume the final form 
\bea
{\cal K}_0(q) =\Delta(q)F^2(q)X_0^{[1]}(q) \,. 
\label{fkern}
\eea

Before solving the system formed by Eqs.(\ref{dirac}) and (\ref{scalar}) 
with ${\cal K}_0(q)$ given by Eq.~(\ref{fkern}), it is instructive 
to study the numerical impact that each of the functions composing ${\cal K}_0(q)$
has on the value of the resulting quark mass.

The results of this exercise are presented in Fig.~\ref{fig3}, where  on the left 
panel we show the support that ${\cal K}_0(q)$ receives when we turn on 
one by one the Green's functions that
compose it. Without a doubt, the biggest numerical contribution 
comes from the ghost dressing function, $F(q^2)$. Nonetheless, one should not 
underestimate the effect caused by the 
scattering kernel, $X_0^{[1]}(q)$, which is responsible for a considerable contribution to the
dynamical mass generation, as presented on the right panel of Fig.~\ref{fig3}.
On this panel, we show the corresponding dynamical quark masses that are obtained 
solving the system of Eqs.(\ref{ker1}) and (\ref{scalar})
when ${\cal K}_0(q)$ assumes one of the four forms presented in the left panel. 
Notice that, neither ${\cal K}_0(q)=\Delta(q)$ nor ${\cal K}_0(q)=\Delta(q)F(q)$
furnish the right amount of support necessary to trigger chiral symmetry breaking.
The chiral symmetry is only broken when ${\cal K}_0(q)=\Delta(q)F^2(q)$ and, phenomenologically compatible values 
are only obtained when the effects of $X_0^{[1]}(q)$  are incorporated in the ${\cal K}_0(q)$. It is important to
mention that, although $X_0^{[1]}(q)$ does not provide a sizable support for ${\cal K}_0(q)$
in the deep IR, the small contribution it furnishes in the intermediate region
is enough for increasing the mass from   \mbox{${\mathcal M}(0)=190$ MeV} to
\mbox{${\mathcal M}(0)=295$ MeV}. This result  
is consistent with previous observations in the literature~\cite{Roberts:1994dr, Maris:1999nt}, 
stating that  the support crucial for quark generation 
originates from the intermediate region of the integration momenta.

Let us now return to 
the solution of the system formed by Eqs.(\ref{dirac}) 
and (\ref{scalar}).   
Substituting  $\Delta(q^2)$, $F(q^2)$, and $X_0^{[1]}(q)$ into  Eqs.(\ref{dirac}) 
and (\ref{scalar}), with the modification of  Eq.~(\ref{fkern}), 
and fixing  the Casimir eigenvalue in the fundamental representation
{\it i.e.} \mbox{$C_{\rm r} = C_{\rm F}=4/3$},  we determine numerically 
the unknown functions $A(p^2)$ and $B(p^2)$.

\begin{figure}[!t]
\begin{minipage}[b]{0.45\linewidth}
\centering
\includegraphics[scale=0.4]{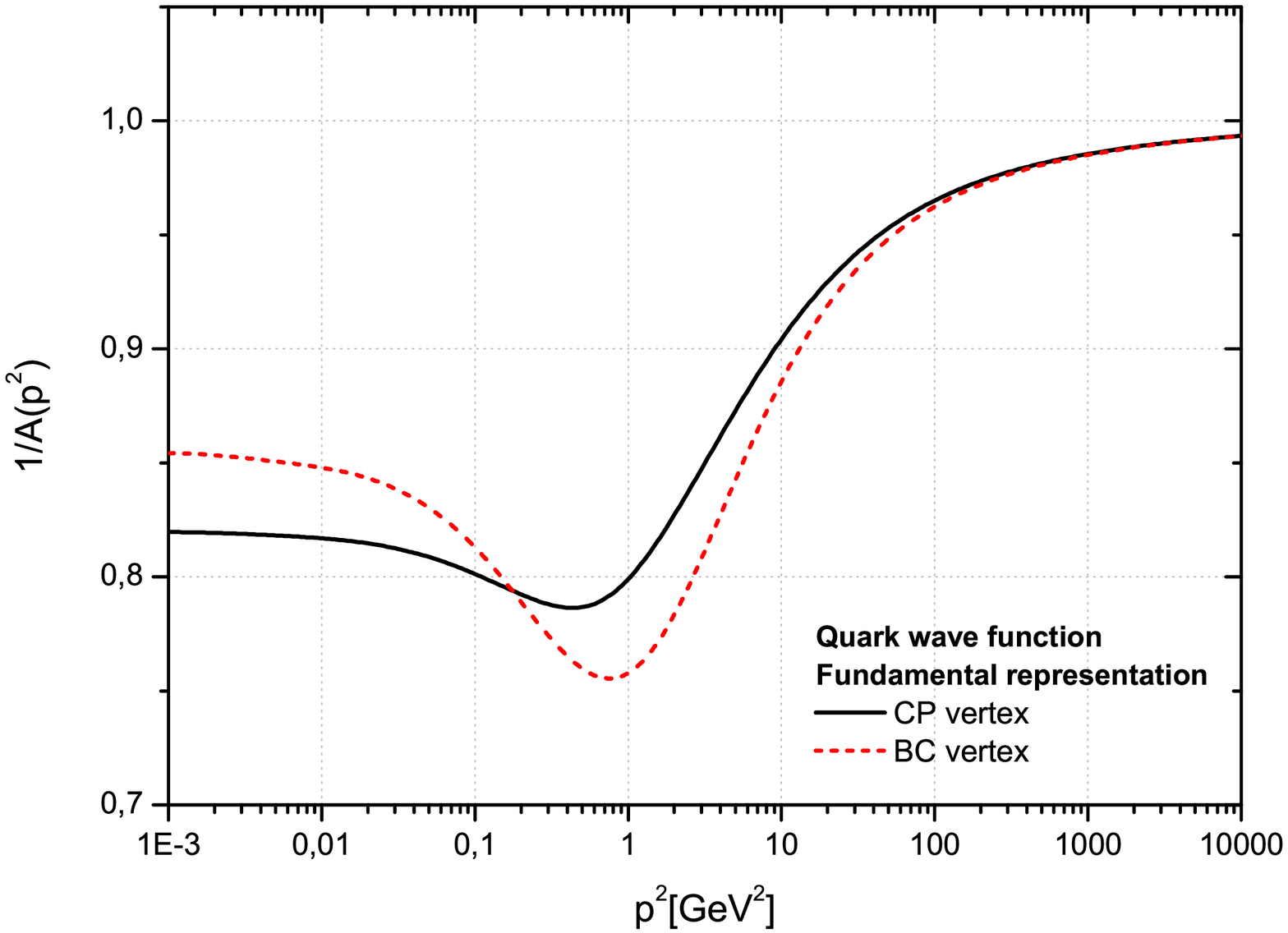}
\end{minipage}
\hspace{0.5cm}
\begin{minipage}[b]{0.50\linewidth}
\centering
\includegraphics[scale=0.4]{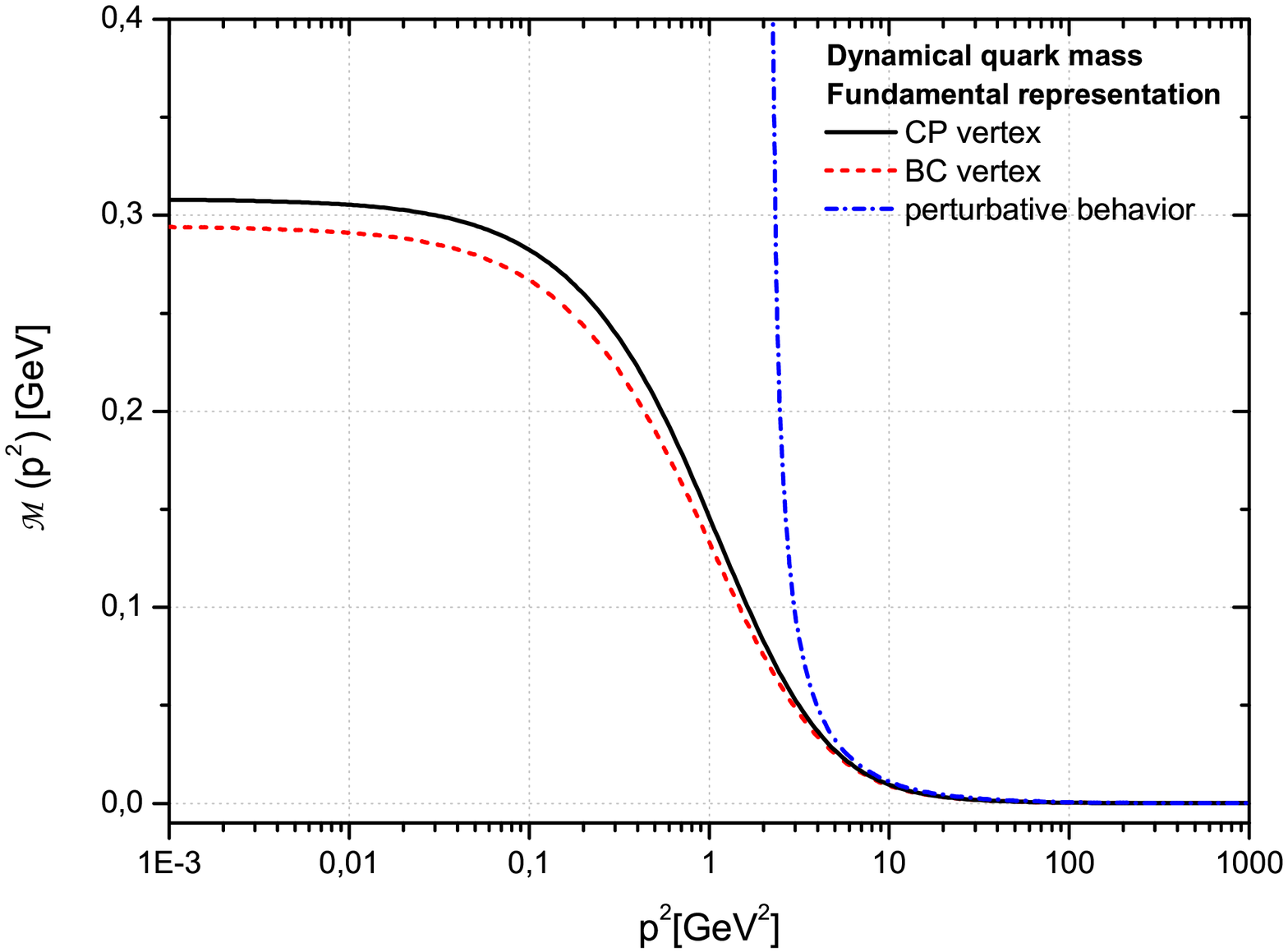}
\end{minipage}
\vspace{-1.0cm}
\caption{{\it Left panel}: The numerical solution for the quark wave function $A^{-1}(p^2)$ in the fundamental 
representation  when 
the non abelian versions of BC (red dashed curve) and CP (black continuous curve) vertices.  
are employed 
{\it Right panel}: The numerical solution for the  dynamical quark mass ${\mathcal M}(p^2)$. The
 red dashed curve represents the dynamical mass generated when the BC vertex is used, while the black continuous line
is the solution found with the CP vertex.}
\label{fig4}
\end{figure}

On the left panel of Fig.~\ref{fig4}, the red dashed line represents  the numerical result 
for the quark wave function $A^{-1}(p^2)$. As we can see, 
for large values of $p^2$, the function $A(p^2)$ goes to $1$, in agreement
with the discussion presented in the Section \ref{ss4}. In the opposite limit,
$A^{-1}(p^2)$ develops a plateau  saturating in a finite value around $0.85$. 

The red dashed line in the right panel of Fig.~\ref{fig4} represents
the corresponding dynamical  quark mass ${\mathcal M}(p^2)$, obtained as 
the ratio  $B(p^2)/A(p^2)$.
One clearly sees that ${\mathcal M}(p^2)$ 
freezes out and acquires a finite value in the IR, while in the UV it shows the expected 
perturbative behavior given by  Eq.~(\ref{asy}) and represented
by the blue dotted-dashed curve.

For the sake of comparison, we also solve numerically
the coupled system when the non-abelian version of the CP vertex
is employed. As we can see from the continuous black curve of Fig.~\ref{fig4}, the qualitative
behavior of the results obtained with the CP vertex is very similar to that
obtained with the BC vertex. Note that both vertices generate phenomenologically acceptable values for ${\mathcal M}(0)$. More
specifically, for the BC vertex we obtain \mbox{${\mathcal M}(0) = 294 $ MeV} while for the CP vertex the
value is slightly higher  \mbox{${\mathcal M}(0) = 307 $ MeV}.

Once the behavior of the dynamical quark mass is determined, one may attempt  
to reproduce some of the phenomenological parameters that depend  
directly on it. Such a 
parameters is the pion decay constant $f_{\pi}$, which measures the ``strength'' of the CSB. The pion 
decay constant is  defined through the axial-vector transition amplitude for an 
on-shell pion, $\left\langle 0\left|A_5^{a\mu}(0)\right|\pi^{b}(k)\right\rangle = if_{\pi}k^{\mu}\delta^{ab}$. 
Making use of the method developed by Pagels and Stokar~\cite{Pagels:1979hd},  
and Cornwall~\cite{Cornwall:1980zw}, $\bar{f_{\pi}}$ can be expressed in terms of 
the dynamical quark mass as  
\bea
\bar{f_{\pi}^2} &=& \frac{3}{4\pi^2} \int_{0}^{\infty} \!\!\!dy  \frac{{\mathcal M}(y)}{[y+{\mathcal M}^2(y)]^2} \left[
{\mathcal M}(y) -\frac{y}{2}\frac{d{\mathcal M}(y)}{dy}\right]\,.
\label{fpi} 
\eea

Substituting in the above equation the numerical solutions for \mbox{$\mathcal{M}(p^2)$} presented in the  Fig.~\ref{fig4}, we obtain \mbox{$\bar{f_{\pi}}=64.3$ MeV} 
for the  case of the BC vertex, while with the  CP vertex we get \mbox{$\bar{f_{\pi}}= 68$ MeV}. These
values should be compared to the experimental value \mbox{$f_{\pi}= 93$ MeV}~\cite{Nakamura:2010zzi}. 
Evidently, the obtained values 
underestimate $\bar{f_{\pi}}$ by almost $30\%$. The origin of this suppression could possibly be traced back 
to some of the approximation used for the quark-quark-pion proper vertex~\cite{Cornwall:1980zw}. 
Some improved versions of Eq.~(\ref{fpi}) can be found in the literature;  in particular, we will 
employ the expression of~\cite{Barducci:1997jh}, where a correction term is added to Eq.~(\ref{fpi}). More
specifically, 
\bea
f_{\pi}^2 &=&  \bar{f_{\pi}^2} +  \delta f_{\pi}^2 \,,
\label{fpi_gatto} 
\eea
with
\be
\delta f_{\pi}^{2} = \frac{3}{4\pi^2} \int_{0}^{\infty} \!\!\!dy \left[  y^3 \left(\frac{d{\mathcal M}(y)}{dy}\right)^2 - y^2 {\mathcal M}^2(y)\left(\frac{d{\mathcal M}^2(y)}{dy}\right)
-y^2 {\mathcal M}^2(y)\frac{d{\mathcal M}(y)}{dy} \right]\,.
\label{correction}
\ee
Adding this term  to the expression of Eq.~(\ref{fpi}), we obtain \mbox{$f_{\pi}= 76.4$ MeV}
for the case of the BC vertex, and \mbox{$f_{\pi}= 80.6$ MeV} for the  CP vertex. 
Although the results are still below the experimental value, 
the correction added clearly contributes in the right direction.

Finally, we will compute  the quark condensate, which plays the role of the order parameter for dynamical CSB. The
quark condensate at scale of $\nu=1\,\mbox{GeV}^2$ is given by~\cite{Roberts:1994dr}
\be
\left\langle \bar{q}q\right\rangle(1\,\mbox{GeV}^2) = -\frac{3}{4\pi^2} 
\int_{0}^{\nu} \!\!\!dy  \frac{y{\mathcal M}(y)}{A(y)[y+{\mathcal M}^2(y)]} \,.
\label{cond} 
\ee

Substituting again the  solution  $ \mbox{$\mathcal{M}(p^2)$}$ and $A(p^2)$ presented in Fig.~\ref{fig4} into 
Eq.~(\ref{cond}), we obtain \mbox{$\left\langle \bar{q}q\right\rangle (1\,\mbox{GeV}^2)=\,(211 \mbox{MeV})^3$} when 
the BC vertex is employed and  \mbox{$\left\langle \bar{q}q\right\rangle (1\,\mbox{GeV}^2)=\,(217 \mbox{MeV})^3$}. 
This value should be compared to the typical value of the quark condensate  
\mbox{$\left\langle \bar{q}q\right\rangle (1\,\mbox{GeV}^2)=\,(229 \pm 9 \mbox{MeV})^3$}~\cite{Gasser:1982ap}.

\subsection{Chiral symmetry breaking in the adjoint}

Next we will  solve the same system of integral equations formed by Eqs.(\ref{dirac}) 
and (\ref{scalar}) for both BC and CP kernels, given by Eqs.~(\ref{kernels}) and (\ref{kcp}) respectively,
when the Casimir eigenvalue in the adjoint representation
{\it i.e.} $C_{\rm r} = C_{\rm A}=3$. 

When one switches from the fundamental 
to the adjoint representation, the overall effect in the gap equation is an enhancement factor of 9/4 
due to the difference in the corresponding Casimir eigenvalues. 
Of course, due to the  nonlinear nature of the gap equation, the 
wave function $A^{-1}_{\rm adj}(p^2)$ and the adjoint mass $\mathcal{M}_{\rm adj}(p^2)$ 
are not obtained from their quark counterparts through simple multiplication 
by 9/4.

\begin{figure}[!t]
\begin{minipage}[b]{0.45\linewidth}
\centering
\includegraphics[scale=0.4]{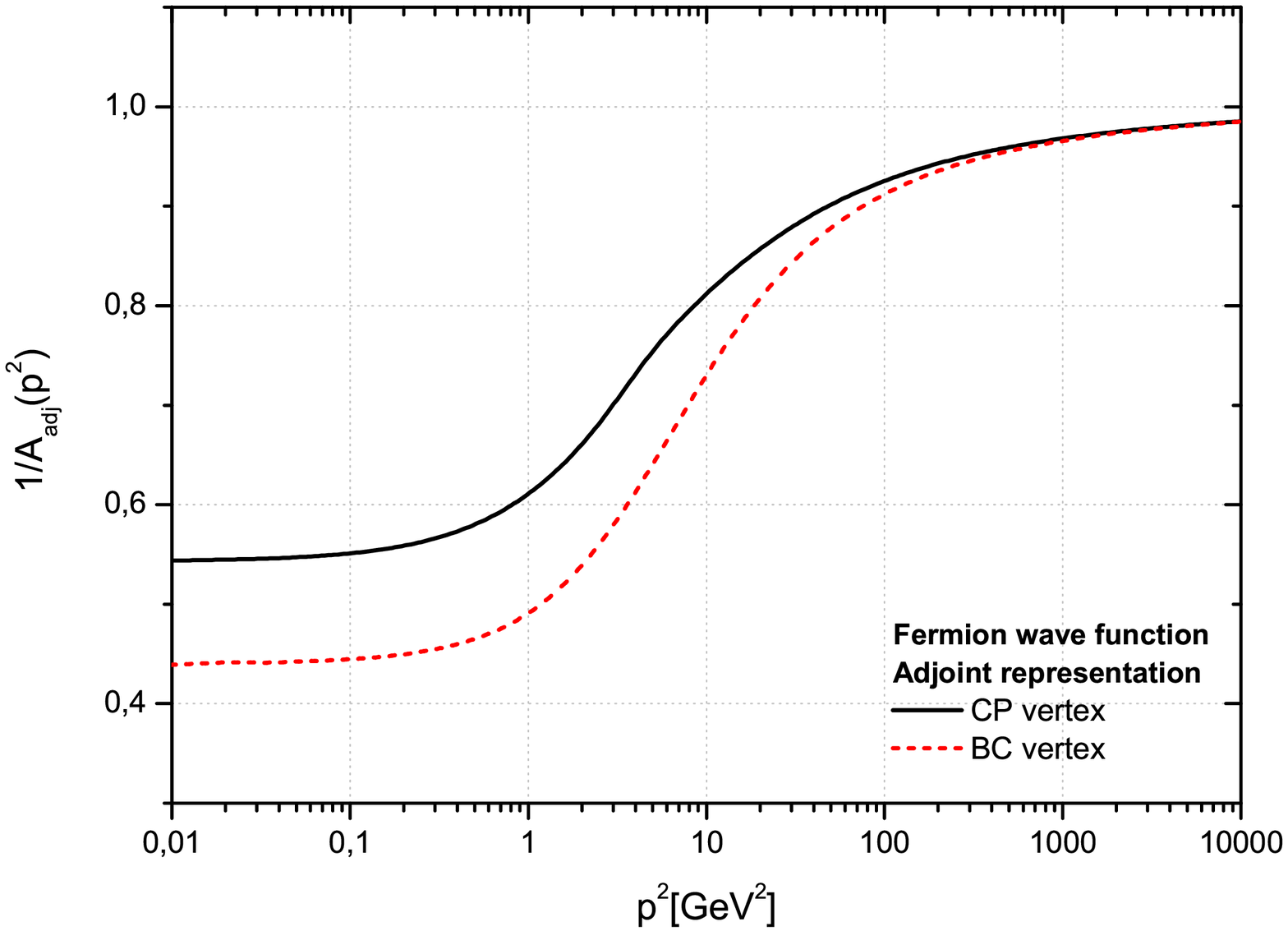}
\end{minipage}
\hspace{0.5cm}
\begin{minipage}[b]{0.50\linewidth}
\centering
\includegraphics[scale=0.4]{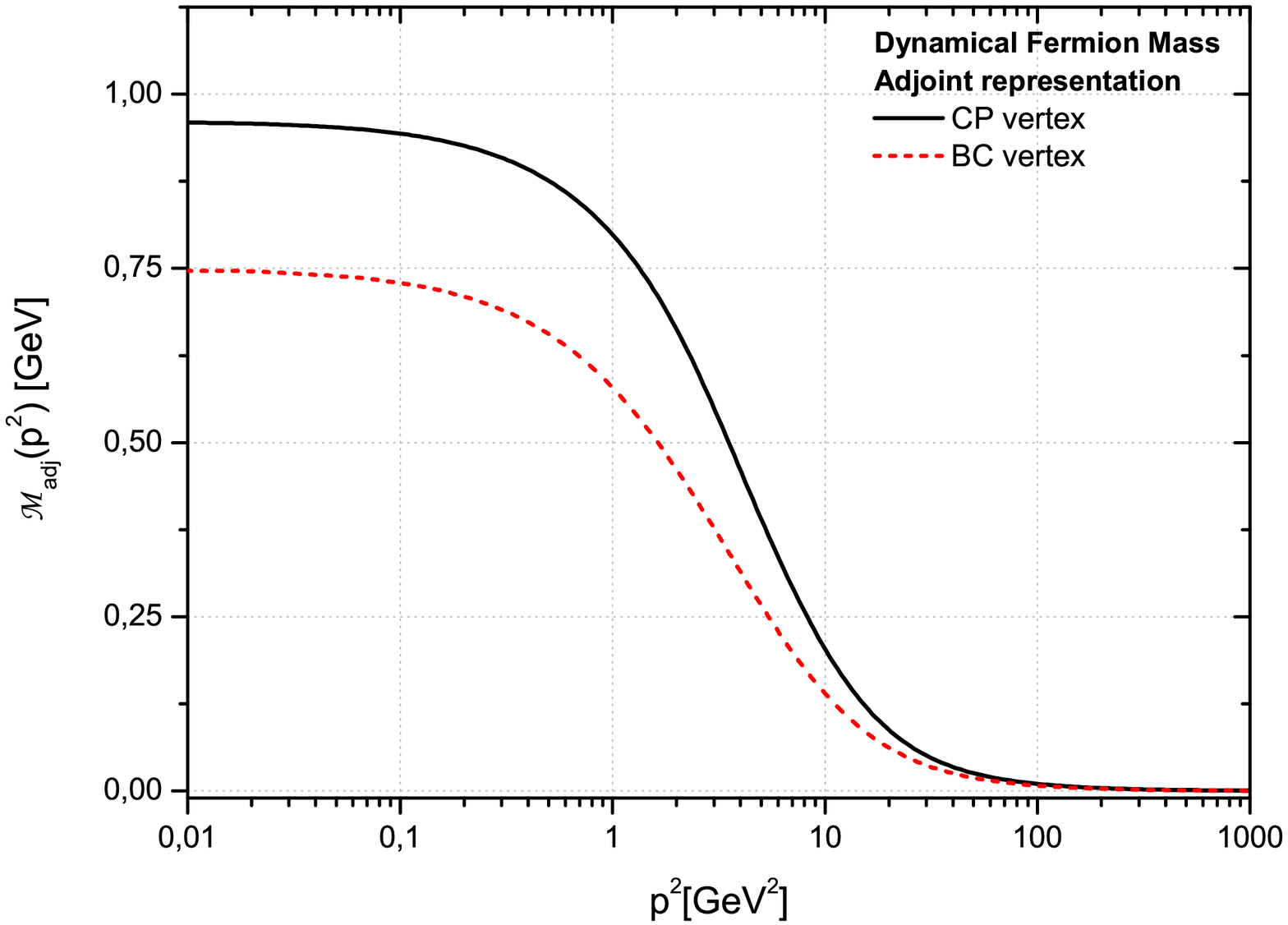}
\end{minipage}
\vspace{-1.0cm}
\caption{{\it Left panel}: The numerical solution for the fermion wave function $A^{-1}_{\rm adj}(p^2)$ in the 
adjoint representation  when 
the modified BC vertex (dotted red curve) and the modified CP vertex (black continuous curve)   
are employed. 
{\it Right panel}: The numerical solution for the  dynamical fermion mass ${\mathcal M}_{\rm adj}(p^2)$. The
dotted red curve represents the dynamical mass generated when the BC vertex is used. The black continuous line
is the solution obtained with the CP vertex.}
\label{fig5}
\end{figure}

The numerical results for the adjoint representation 
are shown in the Fig.~\ref{fig5}. On  the left panel, we compare  
the fermion wave functions, $A^{-1}_{\rm adj}(p^2)$, when we use the modified BC and CP vertices.
As expected,  both solutions displays the right asymptotic behavior while
in the IR limit the fermion wave functions saturate in smaller values compared
to those of the fundamental  representation. In addition, we can notice
that in the adjoint,  $A^{-1}_{\rm adj}(p^2)$ does not display anymore a minimum
as it does in the fundamental representation.

On the right panel, we show the fermion dynamical mass $\mathcal{M}_{\rm adj}(p^2)$.
We see that the infrared saturation of  $\mathcal{M}_{\rm adj}(p^2)$ occurs
for higher values compared to the  values of $\mathcal{M}(p^2)$ in the fundamental.
In particular, when the modified BC vertex is employed, 
one obtains $\mathcal{M}_{\rm adj}(0)=750 $ MeV, while for the CP vertex $\mathcal{M}_{\rm adj}(0) = 962$ MeV.
Clearly we can see hat the latter values are higher than $9/4\mathcal{M}(0)$. 
In addition, it  is interesting to  notice that the adjoint representation is more sensitive to the 
change from the BC to the CP vertex, since
the difference between the results obtained with the both vertices is much more pronounced 
here than in the fundamental representation. 

\section{Discussion and Conclusions}
\label{concl}

In this article we have studied the CSB pattern that emerges 
when the gap equation is combined with the available lattice 
data for the gluon and ghost propagators~\cite{Bogolubsky:2007ud}.  
Particular attention has been payed to the way the ghost sector 
enters into the gap equation. In particular, a complete 
Ansatz for the longitudinal part of the quark-gluon vertex has been constructed,  
which, in addition to the ghost-dressing function~\cite{Fischer:2003rp}, 
captures the full dependence on the various form factors 
composing the quark-ghost scattering kernel.  
This new vertex has been used to derive a more complete version of the 
gap equation, containing additional contributions  
from the quark-ghost scattering kernel. This latter quantity satisfies its own dynamical 
equation, which is coupled to the gap equation in a complicated way. 

In order to reduce the complexity of the problem, we have 
restricted ourselves to the interplay of the gap equation with the 
only the scalar part, $X_0$, of the quark-ghost scattering kernel, deriving its  
``one-loop dressed'' expression. The form factor $X_0$ was then determined from 
this latter expression, for a special momentum configuration, using the lattice data 
for the gluon and ghost propagators appearing in it. 
The resulting expression for $X_0$, shown in Fig.~\ref{sk}, 
reaches its maximum at a momentum of a few hundred MeV, 
where it displays a 25$\%$ enhancement compared to its tree-level value. 
We emphasize that the numerical effect of including $X_0$
into the gap equation is rather sizable; indeed, as can be seen 
from Fig.~\ref{fig3},  it accounts for about 30$\%$ of the final result for the quark mass.
 
Finally, all ingredients are substituted into the gap equation, which is solved for the 
case of quarks (fundamental representation), and for fermions in the adjoint representation.
The quark mass obtained is about 300 MeV, in good agreement with phenomenology, 
and is rather insensitive to the details of the quark-gluon vertex employed (BC vs CP).
The corresponding mass obtained for the  adjoint fermions displays a stronger dependence on the 
form of this vertex, varying between (750-962) MeV.

It   is  important   to   emphasize  that   there   is  a crucial  
complementarity  between using  lattice  data as  input  into the  gap
equation, and,  at the same time,  employing an Ansatz  for the vertex
that  captures suitably the dependence  on quantities  such  as the
ghost-dressing  function   and  the  quark-ghost   scattering  kernel.
Indeed, the  additional dependence on  the ghost sector  stemming from
the vertex  would be insufficient for  getting the correct  CSB if one
were  not  to  use  the  lattice  results,  which  capture  fully  the
underlying dynamics, displaying a sizable enhancement with respect to
other non-perturbative approaches. Similarly, no realistic CSB pattern
can be obtained if one were to substitute the lattice ingredients into
the gap  equation obtained from  a less elaborate  quark-gluon vertex,
i.e., one  that fails  to include the  effects due to  the quark-ghost
scattering  kernel. Thus, at least  within  our framework,  it is  the
interplay between  these two points that finally  provides an adequate
description of CSB.
 
Amplifying on the previous point, let us mention that 
the SDE analysis presented in~\cite{Aguilar:2008xm}, even though 
it reproduces qualitatively the lattice results, 
and most importantly accounts for the observed IR finiteness of 
the gluon propagator and the ghost-dressing function, it   
underestimates the size of both quantities by a significant amount. 
Clearly, it is an important challenge 
for the SDE approach of the PT-BFM formalism~\cite{Aguilar:2006gr,Aguilar:2008xm,Binosi:2009qm}   
to  eliminate the aforementioned quantitative discrepancy.
Perhaps the most obvious step would be to 
extend the analysis of~\cite{Aguilar:2008xm}
beyond the ``one-loop dressed'' approximation, 
given that there is no a-priori guarantee that the omitted ``two-loop dressed''   
contributions are numerically depreciable. Needless to say, from the technical point 
of view, such an attempt would constitute a formidable task. 

Another possible source of enhancement 
for the Green's functions in question 
may be related to the non-trivial structure 
of the vacuum, and in particular with the presence of solitonic structures, such as vortices 
or monopoles. These classic field configurations are closely linked to the mechanisms of 
confinement~\cite{Greensite:2003xf} 
and CSB~\cite{de Forcrand:1999ms}, 
and are known to affect the shape and size of the fundamental Green's 
functions of the theory~\cite{Gattnar:2004bf}.
A particularly instructive example of how  
to include such effects at the level of the SDEs
has been presented in~\cite{Szczepaniak:2010fe}.

\appendix

\section{\label{app1} The gap equation in the PT-BFM language}

Given the recent reformulation of the SD series within the PT-BFM 
framework~\cite{Aguilar:2006gr,Binosi:2008qk,Binosi:2009qm},  it is conceptually interesting 
to re-express the gap equation using the PT-BFM terminology.
Let us stress from the beginning, however, that, unlike what happens 
in the case of the gluon self-energy~\cite{Binosi:2009qm}, where the corresponding SD equation in the PT-BFM formalism 
is vastly different from the conventional one, the 
gap equations obtained within both formalisms are completely identical.  

Let us start the discussion by pointing out that the 
PT quark-self energy coincides 
with the conventional one in the Feynman gauge,
both perturbatively (to all orders), as well as 
nonperturbatively.
The reason has been explained in detail in the 
related literature; here it should suffice to mention 
that in the Feynman gauge there are no pinching momenta, 
and all three-gluon vertices appearing in the quark-self energy are ``internal'', 
in the sense that all legs are irrigated by 
virtual momenta, and, therefore, they should not undergo the standard PT decomposition, 
a key ingredient in the construction of the PT gluon self-energy.
An equivalent way of saying this in the BFM language~\cite{Abbott:1980hw} 
is that, unlike gauge fields, fermionic fields 
are not split into a background and a quantum component. 
Therefore, the BFM fermion propagators are the same as the 
conventional ones (in all gauges). 
However, given that all nonperturbative ingredients we will use 
come from lattice simulations in the Landau gauge, the gap equation we study here 
is {\it not} the genuine PT gap equation.
Away from the Feynman gauge, one must  
switch to the BFM language, or, equivalently, to the 
generalized PT. In any case, the central result remains the same:
the gap equation in the Landau gauge is the same as the conventional one.

The PT-BFM gluon self-energy, denoted by 
$\widehat\Delta$, behaves in many aspects as that of QED;
in particular, the product $g^2\widehat\Delta$ is RG-invariant 
(for any choice of the BFM gauge-fixing parameter).
Of course,  the propagator appearing in Eq.~(\ref{rsenergy1}) is {\it not} $\widehat\Delta$ 
but rather the conventional $\Delta$; indeed, background field propagators do 
not propagate inside loops, only quantum ones. 
However, there exists a set of powerful identities
that allows one to establish some important 
connections~\cite{Grassi:1999tp,Binosi:2002ez,Aguilar:2009nf}. Specifically,  
\bea
\Delta(q^2) &=& 
\left[1+ G(q^2)\right]^2 \widehat{\Delta}(q^2)\,,
\nonumber\\
F^{-1}(q^2) &=& 1+G(q^2)+L(q^2)\,.
\label{BQIs}
\eea
The functions $G(q^2)$ and $L(q^2)$ are the two form factors of 
a particular two-point function, denoted by $\Lambda_{\mu\nu}(q)$, defined as~\cite{Binosi:2009qm,Aguilar:2009pp} 
\bea
\Lambda_{\mu\nu}(q) &=& -ig^2C_A
\int_k H^{(0)}_{\mu\rho}
D(k+q)\Delta^{\rho\sigma}(k)\, H_{\sigma\nu}(k,q)
\nonumber\\
&=& g_{\mu\nu} G(q^2) + \frac{q_{\mu}q_{\nu}}{q^2}L(q^2) \,.
\label{LDec}
\eea

Turns out that the function $L(q^2)$ is subleading both in the IR and in the UV; therefore, 
for the purposes of this argument can be safely neglected~\cite{Aguilar:2009nf}. Then, the combination of the 
two identities in (\ref{BQIs}) leads to the approximate relation 
\be
\widehat{\Delta}(q^2) = \Delta(q^2) F^{2}(q^2)\,,
\label{bqiapp}
\ee
leading to exactly the same conclusion as before.  

Finally, the above discussion may be recast in the more intuitive 
language of an effective (running charge),
traditionally employed in QED-inspired studies of QCD. 
From the (dimensionful) RG-invariant quantity $g^2\widehat\Delta$ introduced above, one may 
define a nonperturbative effective charge, denoted by $\alpha(q^2)$, as    
\be
g^2\widehat\Delta(q^2) = \frac{4\pi\alpha(q^2)}{q^2+m^2(q^2)} \,,
\label{effch}
\ee
where $m^2(q^2)$ is the momentum-dependent dynamical gluon mass, whose value in the deep IR 
is about $(500-700)$ MeV. Using Eq.~(\ref{bqiapp}), and after implementing Eq.~(\ref{rgi_improve}), 
the alternative (and completely equivalent) form 
of Eq.~(\ref{dirac}) reads~\cite{Papavassiliou:1991hx,Haeri:1990yj} (setting $q\equiv p-k$)
\be
A(p^2)= 1 + 4\pi C_{r}  \,\int_{k}\, \frac{\alpha(q^2)}{q^2+m^2(q^2)} 
\frac{{\cal K}_A(k,p)}{A^2(k^2)k^2+B^2(k^2)}\,, 
\label{dirac2}
\ee
with an exactly analogous expression for $B(p^2)$. 
Note that finally there is no explicit reference to $F(q^2)$, because it has 
all been absorbed into the definition of the effective charge  $\alpha(q^2)$.
The RG-invariance of this equation can be easily established, given that both $\alpha(q^2)$, 
the gluon mass  $m^2(q^2)$, are RG-invariant. The final inclusion of the $X_0^{[1]}$ into 
 Eq.~(\ref{dirac2}) is straightforward.

\section*{Acknowledgments}
The research of J.~P. is supported 
by the European FEDER and  Spanish MICINN under grant FPA2008-02878, 
and the Fundaci\'on General of the UV. The work of  A.C.A  is supported by the Brazilian
Funding Agency CNPq under the grant 305850/2009-1.

\end{document}